\documentclass{article}

\begin{document}

\def\square{\,\hbox{\vrule\vbox{\hrule\phantom{N}\hrule}\vrule}\,}

\def\bfxi{\mbox{\boldmath $\xi$}}

\def\bfvartheta{\mbox{\boldmath $\vartheta$}}

\thispagestyle{plain}
\def\upcirc#1{\vbox{\ialign{##cr
$\circ\!$cr \noalign{\kern-0.1 pt\nointerlineskip}
$\hfil\displaystyle{#1}\hfil$cr}}}
\def\X{\bf X}
\def\thorn{\hbox{\rm I}\kern-0.32em\raise0.35ex\hbox{\it o}}
\def\edth{\hbox{$\partial$\kern-0.25em\raise0.6ex\hbox{\rm\char'40}}}
\def\edthbar{\overline{\edth}}
\def\taub{\overline{\tau}}
\def\obar{\overline{o}}
\def\iotabar{\overline{\iota}}
\def\thornprime{\thorn^\prime}
\def\edthprime{\edth^\prime}

\def\i{\iota}
\def\P{{\bf \Psi}}
\def\Thorn{{\bf \thorn}}
\def\Fi{{\bf \Phi}}
\def\bfi{\bf \phi}
\def\bnab{{\bf \nabla}}
\def\L{{\bf \Lambda}}
\def\bxi{{\bf \Xi}}
\def\bep{{\bf \epsilon}}
\def\bdelta{\mbox{\boldmath $\delta$}}
\def\bDelta{{\bf \Delta}}
\def\thornp{{\thorn^\prime}}
\def\edthp{{\edth^\prime}}
\def\bfeta{{\bf \eta}}
\def\pmb#1{\setbox0=\hbox{#1} \kern-.025em\copy0\kern-\wd0
\kern.05em\copy0\kern-\wd0 \kern-.025em\raise.0233em\box0 }
\def\bedth{{\pmb{\edth}}}
\def\bthorn{{\pmb \thorn}}
\def\thornp{{\thorn^\prime}}
\def\edthp{{\edth^\prime}}
\def\bfeta{{\bf \eta}}
\def\bthornp{{\bthorn^\prime}}
\def\bedthp{{\bedth^\prime}}
\def\alemf{{\nabla^{f\plf}\nabla_{f\plf}}}
\def\Ib{\overline{\bf I}}
\def\I{\bf I}
\def\ei{{\rm i}}
\def\Ph{\nthorn}
\def\D{\nedth}
\parindent=0pt
\def\pmb#1{\setbox0=\hbox{#1}  \kern-.025em\copy0\kern-\wd0
 \kern.05em\copy0\kern-\wd0
  \kern-0.025em\raise.0433em\box0 }
\def\half{{\scriptstyle {1 \over 2}}}
\def\third{{\scriptstyle {1 \over 3}}}
\def\quarter{{\scriptstyle {1 \over 4}}}
\def\o{{\pmb o}}
\def\i{{\pmb{$\iota$}}}
\let\gt=\mapsto
\let\la=\lambda
\def\a{\overline a}
\def\daa{\nabla_{AA'}}
\def\ob{{\overline\o}}
\def\ib{{\overline\i}}
\def\T{{\bf T}}
\def\bp{{\bf p}}
\def\bq{{\bf q}}
\def\0{\pmb 0}
\def\1{\pmb 1}
\def\2{\pmb 2}
\def\3{\pmb 3}
\def\bPhi{{\pmb{$\Phi$}}}
\def\bPsi{{\pmb{$\Psi$}}}
\def\>{\phantom{A}}
\def\sym{{\sum_{sym}}}
\def\thorn{\hbox{\rm I}\kern-0.32em\raise0.35ex\hbox{\it o}}
\def\edth{\hbox{$\partial$\kern-0.25em\raise0.6ex\hbox{\rm\char'40}}}
\def\edthbar{\overline{\edth}}
\def\nedth{{\pmb\edth}}
\def\nedthp{{\pmb\edth'}}
\def\thornp{\thorn'}
\def\nthorn{{\pmb\thorn}}
\def\nthornp{{\pmb\thorn'}}
\def\edthp{\edth'}
\def\a{\alpha}
\def\b{\beta}
\def\g{\gamma}
\def\d{\delta}
\def\cd{{\cal D}}
\def\boeta{{\pmb{$\eta$}}}
\def\blambda{{\pmb{$\lambda$}}}
\def\etad{\boeta_{C_1\dots C_N C'_1\dots C'_{N'}}}
\def\p{{\bf p}}
\def\q{{\bf q}}
\def\K{{\bf K}}
\def\R{{\bf R}}
\def\S{{\bf S}}
\def\T{{\bf T}}
\def\I{{\bf I}}
\def\la{\lambda}
\def\lab{\overline\lambda}

\title{Invariant classification and the generalised invariant formalism:
conformally flat pure radiation metrics, with zero cosmological constant.\footnote{This is an expanded version of the publication \cite{BER2}. Also some typos and numerical coefficients have been corrected compared  to the published version.}}
\author {Michael Bradley
\\ Department of Physics,
 \\ Ume\aa \ universitet,\\  Ume\aa\\
Sweden  S-901 87\\
 michael.bradley@physics.umu.se
 \and S. Brian Edgar
\\ Department of Mathematics,
 \\ Link\"{o}pings universitet,\\ Link\"{o}ping\\
Sweden S-581 83\\
bredg@mai.liu.se
\and M.P. Machado Ramos
\\ Departamento de Matem\' atica,
\\ para a Ci\^encia e Tecnologia,
\\ Azur\'em 4800-058 Guimar${\hbox {a}}$es,
\\
Universidade do Minho,
\\ Portugal\\
mpr@mct.uminho.pt
 }

\maketitle

\section*{Abstract.}
Metrics obtained by integrating within the generalised invariant
formalism are  structured around their intrinsic coordinates, and
this considerably simplifies their invariant classification and
symmetry analysis. We illustrate this by presenting a simple
and transparent complete invariant classification of the conformally flat pure
radiation metrics (except plane waves) in such intrinsic
coordinates; in particular we confirm that the three apparently
non-redundant functions of one variable are genuinely non-redundant, and
easily identify the subclasses which admit  a Killing and/or a
homothetic Killing vector. Most of our results agree with the
earlier classification carried out by Skea in the different
Koutras-McIntosh coordinates, which required much more involved
calculations; but there are some subtle differences. Therefore,
 we also rework the classification in the
Koutras-McIntosh coordinates, and by paying attention to some of
the subtleties involving arbitrary functions, we are able to
obtain complete agreement with the results obtained in intrinsic
coordinates. In particular, we have corrected and completed
statements and results by Edgar and Vickers, and by Skea, about
the orders of Cartan invariants at which particular information  becomes
available.

\

{\bf PACS numbers:} \ 0420,\ 1127

\

\section{Introduction}

\subsection{Integration in tetrad formalisms}

The integration procedure \cite{edGHP}, \cite{held1}, \cite{held2}
 developed in the GHP formalism
\cite{ghp}
--- which is built on two intrinsic vectors or spinors  picked out
by the spacetime geometry --- has recently been generalised \cite{edvic}, \cite{edram1},
\cite{edram2}, \cite{edram3}
 to the
GIF (generalised invariant formalism) \cite{maria1}, \cite{maria2}, 
\cite{maria3}--- which is built on one intrinsic vector or spinor   picked out by
the spacetime geometry. Compared to the familiar integration
procedures in NP formalism \cite{np}, \cite{edGHP}, this procedure
is much more efficient and avoids detailed complicated gauge
calculations which have the potential for errors. It also supplies
the metric in a natural form, with coordinates chosen --- as far as
possible --- in an intrinsic and invariant manner, which permits
comparatively simple invariant classification procedure, equivalence
problem analyses and  symmetry investigations.

\subsection{Equivalence problem and invariant  classification of metrics}

The equivalence problem is the problem of determining whether the
metrics of two spacetimes are locally equivalent, and the original
contribution of Cartan \cite{cartan} directed attention to the
Riemann tensor and its covariant derivatives up to $(q+1)$th order,
${\cal R}^{q+1}, $ calculated in a particular frame.

  In going from ${\cal R}^q$ to ${\cal
R}^{q+1}$ for a particular spacetime, if  there is no new functionally independent Cartan scalar
invariant  and  ${\cal R}^q$ and ${\cal R}^{q+1}$ have equal isotropy group, then all the local
information that can be obtained  about the spacetime is contained
in the set  $ {\cal R}^{q+1}$. The set $ {\cal R}^{q+1}$ is
called the {\it Cartan scalar invariants}  and provide the information
for an {\it invariant classification} of the spacetime.

That the $q+1$-derivatives are needed can be understood in the following way. Call the functionally
independent elements $I^{\alpha}$. We then have
\begin{displaymath}
dI^{\alpha}=I^{\alpha}{}_{\vert K}\omega^K.
\end{displaymath}
Here $K=1,2,...,\frac{n(n+1)}{2}$ numbers a basis on $F(M)$ and
$I^{\alpha}{}_{\vert K}\equiv X_K(I^{\alpha})$ where $X_K$ is the
dual basis to $\omega^K$. Note that if a certain $I^{\alpha}$
appears in the $q$th derivative, then $I^{\alpha}{}_{\vert K}$
appears in the $q+1$ derivative. In the case without symmetries the
relation above can be inverted to give all $\omega^K$ and in
particular the 1-forms spanning the tangent plane of $M$,
$\omega^i$, $i=0,1,...n-1$, in terms of the elements in ${\cal
R}^{q+1}$. Hence the metric is obtained as
$ds^2=\eta_{ij}\omega^i\omega^j$. The proof can be extended to the
case with symmetries \cite{cartan}, \cite{karl}.

Two metrics are equivalent and represent the same spacetime if all
their respective Cartan scalar invariants in ${\cal R}^{q+1}$ can be
equated consistently. It is important to note that although there
will be no new information about essential coordinates in the step
from ${\cal R}^{q}$ to ${\cal R}^{q+1}$, there {\it may} be other
new information,  in particular about inconsistencies and also the
nature of apparently non-redundant functions (including constants).

A complete invariant  classification  of a spacetime should include
the true nature of apparently  non-redundant functions, i.e.,
whether they are genuinely non-redundant, or can be transformed away
by a coordinate transformation. A convenient way to do this for a
particular spacetime is to carry out a simplified form of an
equivalence problem:  a metric with coordinates $x^i, i=1,\ldots,
4,$ and non-redundant functions $f^{\alpha}, \alpha = 1,\ldots m$,
can be tested for equivalence against a second metric, which is its
copy   with exactly the same structure, but where we have relabelled
the  coordinates $X^i$ and the non-redundant functions $F^{\alpha}$.
If, and only if, the equivalence problem confirms these two metrics
as completely equivalent, and in particular identifies each of the
apparently non-redundant functions with a Cartan invariant, then we
can conclude that the non-redundant functions are essential and
genuinely non-redundant.  (Non-redundant constants can be treated as
a special case of non-redundant functions, and are understood to be
included in our discussions; but see \cite{Pap} for a different
treatment of non-redundant constants.) Note that when working with
an explicit metric the $I^{\alpha}{}_{\vert K}$,
may appear to  "vanish" since they are explicitly calculated. However, if we want to compare two metrics each with
an apparently non-redundant function, say $f(u)$ and $F(U)$ respectively, we do not only need the relation between them, but also
 between their
derivatives: suppose that we find $F(U)=f(u)$ as the last coordinate relation at order $p$, then
for consistency we  need
\begin{displaymath}
\frac{d F}{d U}=\frac{d F}{d u}\frac{d u}{d U}=\frac{d f}{d u}\frac{d u}{d U}.
\end{displaymath}
When comparing this with the relation between $F_U (=dF/dU)$ and $f_u (= df/du)$, that will appear at order $p+1$,
we can then solve for $\frac{d u} {dU}$.

\subsection{Karlhede algorithm for invariant  classification of metrics}

A  practical method for invariant classification was developed by
Karlhede \cite{karl}, using fixed frames. In this algorithm the
number of functionally independent quantities is kept as small as
possible at each step by putting successively the curvature and its
covariant derivatives into canonical form, and only permitting those
frame changes which preserve the canonical form. In practice, rather
than work directly with the Riemann tensor and its covariant
derivatives, it is convenient to use decompositions of their spinor
equivalents; a minimal set of such spinors has been obtained in
\cite{maccullum}. The frame components of ${\cal R}^{q+1}$  in the
canonical
 frame  fixed by the Karlhede algorithm will be called the {\it Cartan-Karlhede  scalar
invariants}\footnote{In the literature these are usually simply
called the {\it Cartan scalar invariants}, but we wish to
distinguish these invariants found using the Karlhede algorithm from
other invariants. }. Although the Karlhede algorithm is more
efficient than the original procedures proposed by Cartan or Brans \cite{Brans}, it may
need to go as far as to ${\cal R}^7$ \cite{karl}, and  as a
consequence, for some spacetimes, long complicated calculations are
required, which usually need computer support, e.g., using the
programme CLASSI \cite{aman}, or the Maple-based GRTensor  programme
\cite{GR}.

 Subsequently, it has been shown how the Karlhede algorithm  can be exploited
 to determine the structure of the isometry group of the spacetime, as
well as subclasses within the spacetime which have additional
isometries \cite{kamac}; more recently the scheme has been the basis
for an algorithm which determines whether a spacetime admits  a
homothetic Killing vector \cite{ks}.

\subsection{Invariant  classification of metrics using GHP and GIF}

The standard approach when calculating an invariant classification
of a spacetime uses the Karlhede algorithm, exploits a minimal set
of  spinors, and the usual presentation is in the NP \cite{np}
formalism. However, recently,  alternative formalisms have been
shown to be more efficient in the
 performance of invariant classifications for certain classes of spacetimes.

In \cite{cdv1}, \cite{cdv2},  \cite{cdI},  an invariant classification procedure
for  (vacuum and non-vacuum)  Petrov Type D spacetimes
was discussed in the context of the GHP formalism: this formalism is
particularly efficient
  when a pair of
 intrinsic spinors has been identified by the geometry, and
calculations  can be carried out
in a spin- and boost-weighted scalar  formalism, rather than restricting the
spin and boost freedom in an ad hoc manner. The (weighted) invariant {\it GHP Cartan scalars}
 used for the invariant classification are closely related to, but not identical with the
 Cartan-Karlhede scalar invariants used in the Karlhede algorithm: the GHP Cartan scalars are frame derivatives of
  frame components of the Riemann spinor, whereas the Cartan-Karlhede invariant scalars are constructed by
  projecting  the minimal set \cite{maccullum} of covariant derivatives of the Riemannn spinor onto the frame.
   It is also emphasised,
   that for invariant
 classifications using the GIF and GHP
formalism, there is no requirement that the frame be fixed as much as possible at each step, unlike this
requirement which is imposed in the Karlhede algorithm.

In \cite{maria3}, \cite{maria4} an invariant classification
procedure for (vacuum and non-vacuum)  Petrov Type N
spacetimes was discussed in the context of GIF;  this
 formalism is particularly efficient  when one intrinsic spinor has been identified by the
geometry, and calculations for an invariant classification can be
carried out  in a  spin- and boost-weighted and null rotation invariant
spinor formalism, rather than restricting the spin, boost and null
rotation freedom of the frame in an ad hoc manner. Within the GIF the  information
for an invariant classification is carried by the {\it Cartan spinor invariants}.  As soon as a second invariant
spinor has been supplied within the GIF calculations, transfer can be made to the weighted scalar GHP formalism,
and work continued
in the GHP formalism using invariant GHP Cartan scalars, which are scalar versions of the Cartan spinor invariants.

\subsection{CFPR spacetimes (excluding plane waves)}

The subclass of CFPR (conformally flat pure radiation) spacetimes,
which are not plane waves, has provided a very good illustration of
the benefits of the GIF integration procedure \cite{edvic};
furthermore, regarding their classification, Skea has also pointed
out that "this application of the equivalence problem provides a non
trivial didactic application  of the use of the invariant
classifications in practice" \cite{skea1}.

Integrating in the NP formalism, Wils \cite{wils} obtained a metric
(containing one apparently non-redundant function of one coordinate)
which was claimed to represent the whole class of CFPR spacetimes
which were not plane waves; subsequently Koutras \cite{kou} showed
that this was the first metric  from which a new essential base
coordinate was obtained at {\it third}  order of the Cartan scalar
invariants, which means that  its invariant classification formally
requires investigation of {\it fourth}  order  Cartan scalar
invariants. Koutras and McIntosh \cite{km} have given a slightly
more general metric in different coordinates; this form includes
plane waves as well as the Wils metric. When Edgar and Ludwig
\cite{edludL}, \cite{edlud1},
 proposed what appeared to be a
more general metric (containing three apparently non-redundant
functions of one coordinate) to represent all CFPR spacetimes, this
provided an ideal opportunity for a non-trivial application of the
equivalence problem.   (The nontrivial nature of the invariant
classification of this metric was emphasised when Skea's
investigation of it \cite{skea1} revealed a bug in the CLASSI
computer programme which is used to handle the complicated
calculations in the usual NP form; furthermore, when the
Edgar-Ludwig metric was used to test the new Maple based invariant
classification package in GRTensor, problems occurred and the
results of the invariant classification published in \cite{pol} are clearly
in error, as Barnes \cite{barnes} has pointed out.)

Skea \cite{skea1} carried out a detailed  investigation of the
equivalence problem in Koutras-McIntosh coordinates for the 
Edgar-Ludwig and Wils metrics:   this analysis confirmed that  the
Edgar-Ludwig spacetime is a genuine generalisation of the  Wils
spacetime, and
 all information about essential coordinates
was obtained by third order, which means that  the procedure will formally
terminate with fourth  order Cartan scalar invariants; Skea also
found the particular subclass of the Edgar-Ludwig spacetime which
coincided with the Wils spacetime, and as well he investigated the
particular subclass of the Edgar-Ludwig spacetime which permitted
one Killing vector.

Subsequently, Edgar and Vickers \cite{edvic} have rederived the
Edgar-Ludwig spacetime using GIF; the metric obtained is only
slightly different in appearance than the versions in \cite{edludL}
or \cite{edlud1}, but deeper examination confirms a  more natural
and simpler structure, because of the fact that the coordinates have
been chosen in an invariant and intrinsic manner.  A striking
illustration of this is that the Killing vector analysis of this GIF
version of the Edgar-Ludwig metric is trivial \cite{edlud3}.

It is therefore expected that an invariant classification of the version of the metric, obtained by GIF in intrinsic coordinates, will be particularly simple, efficient and transparent.

\subsection{Outline}

The first purpose of this present paper is to demonstrate how the
simpler, more natural structure of the  version of the Edgar-Ludwig
spacetime derived in intrinsic coordinates by a GIF analysis
\cite{edvic}, enables us to get a simpler  and more transparent
picture of its invariant classification;  and in particular to
investigate in full detail the role of the three apparently
non-redundant functions, confirming that they are genuinely
non-redundant. The second purpose is to supply the full details of a
highly non-trivial example of an invariant classification, by
working through  the much more complicated calculations required for
a complete invariant classification in Koutras-McIntosh coordinates
of the Edgar-Ludwig spacetime;  this example highlights a number of
subtleties within the invariant classification procedure, as well as
emphasising again, by comparison,  the much simpler calculations for
the intrinsic coordinate version. The third purpose is  to supply
the full details of a highly non-trivial example of an equivalence
problem, by demonstrating  equivalence between the respective
invariant classifications in the intrinsic coordinates and
Koutras-McIntosh coordinates of the Edgar-Ludwig spacetime.

The invariant classification --- carried out in the intrinsic coordinates and tetrad
derived by the GIF analysis \cite{edvic} ---  is given in detail in
Section 2. This completes and corrects the incomplete discussion by
Edgar and Vickers  of an invariant classification of this GIF
version,  in the last section in \cite{edvic}.

In Section 3 we give a classical invariant Karlhede classification of the
GIF version of the Edgar-Ludwig spacetime in its intrinsic
coordinates using the CLASSI programme. This demonstrates explicitly how the Karlhede algorithm requires a slightly different tetrad from that used in Section 2, although of course the basic results are the same, and the presentation is also simple and transparent.  

Skea's results, \cite{skea1} as quoted above for the Edgar-Ludwig
spacetime, are easily confirmed in Sections 2 and 3, in principle:
in general, all the essential coordinates are  obtained at third
order of the Cartan scalar invariants, which means that the
classification process will formally end at fourth order, where
additional information about the non-redundant functions {\it may}
become available. Further investigation reveals that all three
apparently non-redundant functions in the Edgar-Ludwig spacetime are
genuine non-redundant functions, none of which   can be transformed
away by a coordinate transformation. Furthermore, for the  generic
class, all four essential base coordinates are supplied at third
order, and the fourth order Cartan scalar invariants only repeat
information supplied at a lower level; but for one special subclass
it is found that there are no new essential base coordinates after
the three supplied at second order, and so for this subclass the
algorithm formally ends at third order; and for  a second special
subclass it is found that  all four essential base coordinates are
supplied at third order, but that additional information about one
of the non-redundant functions becomes available at fourth order.
Moreover, this second special subclass which requires  a fourth
order Cartan scalar invariant, does not intersect with the Wils
spacetime; hence the inequivalence  of the Wils and the Edgar-Ludwig
spacetimes is established at third order of the Cartan scalar
invariants.

However, the last conclusion disagrees with some of those in
\cite{skea1}, where it was argued that information from the {\it fourth} order Cartan
scalar invariant $D^4 \Phi_{56'} $ was needed to find all relations
for which the Edgar-Ludwig spacetime reduced to the Wils spacetime.
Therefore, in Section 4, we rework the  calculations in
\cite{skea1}, using CLASSI,  and demonstrate how all relations  are
found already at third order. Incompatabilities might appear at the
next order, but in this case all fourth order relations turn out to
be identically satisfied.

In Section 5, we carry out an
equivalence problem for the two different versions of the metric whose respective Karlhede classifications are in
Sections 3 and 4. We work through the problem in complete detail until we obtain as the final result  the explicit coordinate transformations between the two versions.

In Section 6, we illustrate how simple it is to investigate
Killing vectors and homothetic Killing vectors in the version of
the metric with intrinsic coordinates obtained by a GIF integration, compared with the version in Koutras-McIntosh coordinates.

The paper concludes with a summary and discussion.

\section {Invariant classification of Edgar-Ludwig spacetime in intrinsic coordinates from GIF}

Edgar and Vickers \cite{edvic} have rederived all CFPR spacetimes,
which are not plane waves,  using GIF, obtaining  in coordinates $
t,  \ n, \ a, \ b\ $ the metric
\begin{eqnarray}
g^{ij}=\pmatrix{0 &{-1/ a} & 0&0\cr
{-1/ a} & (-2s(t)+2m(t)a+a^2+b^2)/a & {-n/ a} & {-e(t)/ a}\cr
0&{-n/ a} & -1&0\cr
0&{-e(t) / a}&0&-1}
\label{metric1}
\end{eqnarray}

or equivalently

\begin{eqnarray}
\hbox{d} s^2 &=&\Bigl(a\bigl(2s(t)-2am(t)-a^2-b^2\bigr) -e(t)^2-
n^2\Bigr) \hbox{d} t^2-2a \hbox{d} t \hbox{d} n \nonumber\\
&& + 2 n \hbox{d} t \hbox{d} a + 2 e(t) \hbox{d} t  \hbox{d} b -
\hbox{d} a^2 - \hbox{d} b^2 \label{metric12}
\end{eqnarray}
where $m(t) ,\ e(t),\ s(t)$ are non-redundant functions of the
coordinate $t$; this form  includes the possibility of any of $m(t)$
or $e(t)$ or $s(t)$ being constant. We have the same notation as in
\cite{edvic}, except that we replace  $M(t) ,\ E(t),\ S(t)$ with
respectively $m(t)$,\ $e(t)$,\ $s(t)$. This form represents the most
general  metrics for CFPR spacetimes (with zero cosmological
constant); alternative equivalent forms are given in \cite{edludL},
\cite{edlud1}. In \cite{edvic} the coordinate $a$ was defined as a {\it positive} quantity by  $a=\sqrt {\tau \bar \tau}$: 
however, it is easy to see  that such a restriction is simply an artifact of that method, 
and can be removed;
on the other hand, the restriction  $a\ne 0$ is still valid.

In the final section of \cite{edvic} explicit expressions for some GIF
spinor versions of the Cartan invariants were given, and there was
some discussion about  the invariant classification for this
spacetime in the context of these  Cartan spinor invariants. We
shall now complete these tables and discussion. We begin by
repeating the zeroth, first and second order invariants quoted in
\cite{edvic}.

\smallskip

\underbar{At zeroth order,} there is only the one Cartan spinor invariant

\begin{eqnarray}
\Phi = {q^2\over a}
\label{K0}
\end{eqnarray}

\smallskip

\underbar{At first order,} there are four Cartan spinor invariants
\begin{eqnarray}
\Ph \Phi  = \  0, \qquad
\D \Phi  =  {p q^2\over a^2}, \qquad
\D' \Phi  =  {\bar p q^2\over a^2}, \qquad
\Ph' \Phi  = - { q^2\over a^2}\Bigl({qn\over
a} + 3p{\bf \i}+3\bar p \bar {\bf \i} \Bigr)
\label{K1}
\end{eqnarray}
Here we are following GIF notation and conventions in \cite{maria1} and \cite{maria2};
in particular we are using the index-free "new compacted formalism"
outlined in \cite{maria1} and explained more fully in \cite{maria2}:
${\bf \i}$ is a second spinor which is generated in the GIF
analysis, $p$ and $q$ are weighted scalar invariants which represent
the spin and boost freedom; $q$ is real while complex $p$ satisfies
$p\bar p=1/2$. This is the same notation as in \cite{edvic} except
that we have replaced  ${\bf I}, \ P, \ Q, $  with ${\bf \i}, \ p, \
q$ respectively.

It is easy to see that we  may invert (\ref{K0}) and  (\ref{K1})
for $a$, $p$ and $q$ in terms of Cartan spinor invariants, and
also for $( p\,{\bf \i} +\bar{p}\,\overline{\bf{\i}})$; therefore ${\bf \i}$ is not
uniquely determined, (and so neither is $n$) and --- at this level
--- there clearly would remain the gauge freedom of a one parameter
subgroup of null rotations. Since new information about the
essential coordinates has arisen, we must go to the next order.

\smallskip

\underbar{At second order,} a complete set of independent Cartan
spinor invariants is\footnote{Note the typo corrections in the
second and last equations in (\ref{K2}).}
\begin{eqnarray} \Ph \D \Phi  & =  & \ 0, \qquad
\D \D \Phi  =  {2 p^2q^2\over a^3}, \qquad
\D'  \D \Phi  = { q^2\over a^3}, \qquad
\Ph \Ph'  \Phi  = \  0\nonumber \\
\D \Ph' \Phi & = & -{ pq^2\over a^3}\Bigl({3qn\over a} + 6p{\bf \i}+8\bar p \bar {\bf \i}\Bigr)
\nonumber \\
\Ph'\Ph' \Phi & = & { q^4\over a^4}\Bigl(-s(t)-2am(t) -{5\over2}a^2
+{1\over2}b^2+{3n^2\over a}\Bigr) \nonumber \\ && + {3 q^2\over
a^3}\Bigl(4p^2{\bf \i}^2+5p\bar p {\bf \i}\bar {\bf \i}
+\bar{p}^2\bar{\bf \i}^2\Bigr) +12{ q^3n\over a^4}(p{\bf \i}+\bar p
\bar{\bf \i}) \label{K2}
\end{eqnarray}
together with complex conjugates. The GIF commutator equations
enable us to concentrate on this reduced list of independent
invariants.

We can now invert these equations and obtain explicit expressions
for the spinor ${\i}$, as well as for $n$ and the scalar combination
$\Bigl(-s(t)-2am(t) +{1\over2}b^2\Bigr)$, in terms of Cartan
invariants. Thus, at second order, if we make this choice of ${\bf
\i}$ as our second spinor, we will have  fixed the frame completely
(there is no isotropy freedom remaining),  and we also have
determined three essential base coordinates. Moreover, making this
choice of ${\i}$ as the second dyad spinor enables us to transfer to
the simpler GHP formalism (see \cite{edvic} for a fuller discussion
on when this is possible) since we require only the scalar parts of
the remaining non-trivial Cartan spinor invariants, i.e., the GHP
Cartan invariants. 
So we replace (\ref{K2}) with
\begin{eqnarray}
\edth \edth \Phi & = &{2 p^2q^2\over a^3}, \qquad
\edth'  \edth \Phi  = { q^2\over a^3}, \qquad
\edth \thorn' \Phi  = -{ pq^2\over a^3}\Bigl({3qn\over a} \Bigr)
\nonumber \\
\thorn'\thorn' \Phi & = & { q^4\over a^4}\Bigl(-s(t)+am(t)
+{1\over2}a^2 +{1\over2}b^2+{3n^2\over a}\Bigr)     \, .  \label{K2GHP}
\end{eqnarray}
(The easiest way to obtain this table is to operate with the GHP scalars on the GHP version of (\ref{K1}).\footnote{The numerical coefficients in the GHP scalar invariants at this order and higher orders have a few corrections compared to the published version \cite{BER2}: however, these details do not effect any arguments.})
Again --- since we have obtained new information about essential
coordinates
---  we need to go one step further.

\smallskip

\underbar{At third order}, a comparison of the second order
expressions with the zeroth and first order ones shows that the
only possibly independent new information will come from the
following GHP Cartan invariants\begin{eqnarray} \edth\edth \thorn' \Phi  &=&  -12{p^2q^2\over a^4} {q n \over a}
 \label{K30GHP}\\
 \edth\thorn' \thorn' \Phi  &=&  {pq^4\over a^5}\Bigl(-4s(t)+3am(t) +a^2 +
 {2}b^2+15{n^2\over a}\Bigr) -\hbox{i}
 {pq^4\over a^4} b  \label{K31GHP}\\
\thorn'\thorn'\thorn' \Phi   &  =  &    {q^5\over a^6 }\Bigl(-a
s'(t)+a^2m'(t)-9a n m(t)+10n s(t) + abe(t)\nonumber\\
&&- 4 a^2n -5nb^2-15 {n^3\over a}\Bigr) \label{K32GHP}
\end{eqnarray}
and complex conjugates; prime $ '$ denotes differentiation
with respect to $t$. The GHP commutator equations reduce the number
of independent invariants.

\smallskip

\underbar{At fourth order}, a comparison of the third order
expressions with the zeroth, first and second order ones shows
that the only possibly new independent information will come from
the operator  $\thorn'$  acting on the scalar ${X}$ defined by
\begin{eqnarray}
{ X}&= &\thorn'\thorn'\thorn' \Phi  -  {q^5\over a^6 }\Bigl(
-9a n m(t)+10n s(t) 
- 4 a^2n -5nb^2-15 {n^3\over a}\Bigr)\nonumber\\
&=&  {q^5\over a^6 }\Bigl(-a
s'(t)+a^2m'(t) + abe(t)\Bigr) \label{K4}
\end{eqnarray}
where we have excluded from ${X}$ all terms whose behaviour under
the operators we already know from lower orders.
This gives
\begin{eqnarray}  \thorn' { X}   
& = &  {q^6\over a^6
}\Bigl(-s''(t)+am''(t)-5{ns'(t)\over a}+4n m'(t) + be'(t)
\nonumber\\
&& - 5{n b\over a}e(t) + e(t)^2\Bigr) \, . \label{K4GHP}
\end{eqnarray}
All other fourth order GHP Cartan invariants can be
easily seen to duplicate information already identified at lower
orders.

\smallskip

We have already noted that by second order the spin and boost
parameters $p,q$, and a second spinor ${\bf \i}$ are all uniquely
determined intrinsically, and so there is no isotropy freedom;
this means that we can concentrate on the essential base
coordinates. We have also noted  that  we can also solve for three
essential base coordinates $a, n, \bigl(b^2-2s(t)-4am(t)\bigr)$ at
second order; but solving for a fourth coordinate is a little
more complicated, since we have to go to third order,  and the details will depend on
the nature of the  functions $s(t), m(t), e(t)$.

So we get the following cases:

\smallskip

{\bf Case (a)}\   $m(t) \ne $ constant\  and/or $s(t) \ne$
constant\

 In this case,   (\ref{K31GHP}) and its complex conjugate can be
solved for $b$, which gives a fourth essential coordinate; hence
the third essential coordinate simplifies to
$\Bigl(s(t)+2am(t)\Bigr)$ (even further,  by combining with the
rest of (\ref{K31GHP}) to obtain $s(t)$ or $m(t)$).

\smallskip

{\bf Case (b)}\  $m(t) = m_0,\  s(t) = s_0, \ e(t) \ne
$ constant\ , where  $ m_0, s_0,$ are  constants

In this case,   (\ref{K31GHP}) and its complex conjugate simply
duplicate the third essential coordinate  $b$, while (\ref{K32GHP}) can be solved for $e(t)$ which gives
a fourth essential base coordinate.

\smallskip

{\bf Case (c)}\   $m(t) = m_0,\  s(t) = s_0, \ e(t) = e_0,\ $
where  $ m_0, s_0, e_0$ are all constants

In this case,   the last equation
in (\ref{K2}) supplies a third essential base coordinate $b^2$.
It is obvious that no new information about  essential coordinates
is obtained at third order, and so in this case, the procedure
clearly terminates at third order.

\smallskip

For the first two {\bf (a)}, {\bf (b)}, of the above cases, new
information about  the essential coordinates was given at third
order, and so  we know that, in principle,  we need to continue to
fourth order where there \underbar{\it may} be further information
available.  However, since all possible information about
essential coordinates has already been obtained (no isotropy
freedom remaining, and four base coordinates identified) clearly
there can be no new information possible about {\it essential
coordinates}, although  there \underbar{\it may} be new
information about the nature of the apparently {\it non-redundant
functions}.

\smallskip

Summing up, we have found that:

\smallskip

$\bullet$  if at least one of the functions $m(t), s(t), e(t)$, is
not constant  then all four essential base coordinates are
obtained from GHP Cartan  invariants at third order, and the procedure
will therefore formally terminate at fourth order.

\smallskip

$\bullet$ when all of the functions $m(t), s(t), e(t)$ are
constants,  then only three essential base coordinates can be
obtained from Cartan  GHP  invariants; these are obtained  at second
order, and the procedure will therefore formally terminate at
third order.

\medskip

\subsection{The three apparently non-redundant functions}

We next  investigate the three apparently non-redundant functions
$m(t), s(t), e(t)$, and in particular their possible redundancy.

As noted in the introduction, a convenient way to study the invariant
classification of a particular spacetime is to carry out an
equivalence problem of the original spacetime with a relabelled
version of itself. However, because of the very simple structure
of this metric and the close  relationship of its coordinates with
its GHP Cartan invariants it has been easy to draw conclusions on its
classification  from a direct examination of its GHP Cartan
invariants. Continuing in this manner, the three apparently non-redundant
functions $m(t), s(t), e(t)$  can each be directly identified with a linear
combination of the three GHP Cartan invariants \
$\thorn'\thorn'\Phi, \ \Re(\edth \thorn'\thorn'\Phi),\
\thorn'\thorn'\thorn'\Phi$,  \ and hence are genuinely non-redundant,
and cannot be transformed away.

The final step in the classification is to note that in {\bf Cases
(a), (b)}, since the coordinate $t$ does not occur by itself (but
only implicitly in the three functions), it will also be necessary
to have identification of the (non-zero) derivatives $m'(t),
s'(t), e'(t)$ with GHP Cartan invariants. In {\bf Case (a)}
this information is given in the third order GHP Cartan
invariant \ $\thorn'\thorn'\thorn'\Phi, $\ whereas in {\bf Case
(b)} this information is given in the fourth order Cartan spinor
invariant $\thorn'{ X}$, \ (equivalently
$\thorn'\thorn'\thorn'\thorn'\Phi$).

This completes the essentials of the classification procedure.

\smallskip
In \cite{skea1}, the classification of the Edgar-Ludwig metric in
Koutras-McIntosh coordinates was carried out in the form of an
equivalence problem with a relabelled version of itself, and this
approach will be reworked in Section 4.  Therefore in order to
compare with this analogous investigation, we will now also treat
this question of the nature of the three functions more formally and
in full detail in the notation of an equivalence problem. In the
usual manner, we consider a second spacetime whose line element is a
direct copy of the first;  in the second spacetime, we label the
coordinates by $T, N, A, B$, the weighted scalars by $P, Q$, the
second spinor by ${\bf I}$   and the non-redundant functions by
$M(T), S(T), E(T)$. Because of the simple structure of the
classification in these coordinates, we can immediately conclude
from discussions above, and the identifications with Cartan
invariants, the following equivalences,
$$ N = n,\ A = a,\ B = b;\qquad P =  p,\ Q =  q;\qquad \hbox{{\bf I}} = {\bf \i}
$$
Furthermore, from (\ref{K31GHP}) and  the last equation in (\ref{K2}),
\begin{eqnarray} S(T)= s(t),\qquad\qquad   M(T)=  m(t) .
\label{S=S}\end{eqnarray}

When we substitute this information into the remaining   third order Cartan invariant $\thorn'\thorn ' \thorn ' \Phi$, we obtain
\begin{eqnarray}
{\hbox{d} S(T)\over \hbox{d} T}
-2a{\hbox{d} M(T)\over \hbox{d} T} + b E(T)=
{\hbox{d} s(t)\over \hbox{d} t}-2a{\hbox{d} m(t)\over \hbox{d} t}+ be(t) \label{K321} \end{eqnarray}

A full analysis of these identities will involve separate consideration of constant and non-constant functions, as in the above cases.
So  taking each case separately:

\smallskip
{\bf Case (a)}   It follows from (\ref{S=S}) that $t= t (T)$, and so from (\ref{K321}),
by separating out the different coordinates,
\begin{eqnarray}{\hbox{d} m(t)\over \hbox{d} t }= {\hbox{d} M(T)\over \hbox{d} T}, \qquad
{\hbox{d}s(t)\over \hbox{d} t }= {\hbox{d} S(T)\over \hbox{d} T}
.\label{dm=dM}
\end{eqnarray}
Therefore,  from (\ref{S=S})  and (\ref{dm=dM}),  it follows that
\begin{eqnarray}{\hbox{d}T\over \hbox{d} t}
= 1,  \qquad\hbox{and hence,} \quad T = t + t_o,  \label{tT}\end{eqnarray}
where $t_0$ is an arbitrary constant; subsequently from (\ref{S=S}) (and (\ref{K321}))  it follows that
  \begin{eqnarray}
 M(t + t_o)=m( t), \  S(t + t_o)= s(t), \ \bigl( E(t + t_o)=e(t) \bigr) .
\end{eqnarray}
(Of course, it is obvious from (\ref{metric12}) that $t$ has the freedom of an arbitrary additive constant.)
Furthermore,
 by direct substitution of all the earlier results for the
two metrics in this case, we find that the equality for the only non-trivial fourth order GHP Cartan  invariant (\ref{K4GHP}) is   identically satisfied.

 \smallskip
{\bf Case (b)} From (\ref{S=S}) the  functions $S(T), M(T)$ must also be constants, genuinely non-redundant,
 and equal to their counterparts,  i.e., $ S_0 = s_0,\  M_0 =  m_0
$.
 In this case, from (\ref{K321}) it follows that  $E(T)=e(t ) \ (\ne e_0)$ and hence $t=t(T)$. Since we
have no other information at third order we must go to fourth order, where, after substituting all the equalities
which we have found between the two metrics into (\ref{K4GHP}), we obtain
\begin{eqnarray}{\hbox{d} E(T)\over \hbox{d} T}={\hbox{d} e(t)\over \hbox{d} t}
\end{eqnarray}but since we also know that $E(T)=e(t ) $, then we can again deduce (\ref{tT}), and hence
 \begin{eqnarray}  E(t + t_0)=e(t)
\end{eqnarray}

 \smallskip
 {\bf Case (c)}\   In this case also, from (\ref{S=S}) the  functions $S(T), M(T)$ must also be constants, and equal to their counterparts,  i.e., $ S_0 = s_0,\  M_0 =  m_0
$;
and substitutions in (\ref{K321}) gives equality for the
remaining constant function $ E_0 = e_0$, so that all three constants,
which have been identified with GHP Cartan invariants,  are genuinely
non-redundant and cannot be transformed away. This completes the consideration of all third order invariants, and we have already noted that for this case we do not need to consider the fourth order Cartan invariants.  So all that remains is to identify the final (in
this case non-essential) coordinates.  To do this we compare the line element (\ref{metric12}) with
its direct copy in $N, A, B, T$ coordinates; when we make the substitutions which we have just identified,
$N = n, A = a , B =b, $ it follows trivially that $\hbox{d} T = \hbox{d} t$, and hence $T = t + t_0$
where $t_0$ is an arbitrary constant.

\smallskip
In all three cases, we have confirmed  that there is complete
compatability at the appropriate order, so that there is no redundancy between the three
non-redundant functions $m(t), s(t), e(t)$, and  none of them can
be transformed away by a coordinate transformation (this is also
true in the special cases of these functions being non-redundant
constants);  hence this spacetime cannot be presented in a form
with less non-redundant functions than these three.

We noted earlier --- for case {\bf (c)} --- that  the equivalence problem
is formally solved at third order, while ---  for cases {\bf (a), (b)} ---
since we obtained information about essential coordinates at third order,
then formally we need to go to fourth order to complete the equivalence problem, where there \underbar{\it may}
be new information about the non-redundant functions..
However, in the above calculations, we see that --- in case {\bf (a)}  --- we have been able to extract all
possible information about the three non-redundant functions without needing to go to fourth order GHP Cartan invariants,
whereas --- in case {\bf (b)} ---  we required one additional
{\it fourth} order GHP Cartan invariant (\ref{K4GHP}).

\medskip

This section amplifies the discussion on the invariant
classification in the last section of \cite{edvic}, and corrects
the brief comments in the last paragraph in that section. The fact
that the two sets of coordinates and the two sets of non-redundant
functions have been shown to be trivially identical by this analysis
is of course due to the intrinsic nature of the coordinates and
tetrad supplied by the GIF procedure.

\subsection {Deducing the Karlhede classification}

As noted in the Introduction, the general invariant classification procedure can be refined by the Karlhede classification algorithm   whereby the frame is fixed as much as possible at each successive order of the algorithm. The classification which has just been performed has not followed that algorithm; in particular it was clear that at first order of GHP Cartan invariants the second dyad spinor  could have been identified up to the freedom of a {\it one} parameter null rotation, but instead we rather chose to identify the second dyad spinor completely at second order of GHP Cartan invariants. Therefore in order to refine the above classification to a Karlhede-type classification we need to make certain modifications.

We first need to rewrite $\i \to  \i + q \bar p n\o /3a$ in the last equation in (\ref{K1}); this ensures that for the scalar part of this Cartan spinor invariant, $\thorn' \Phi = 0$,  which is the condition in the Karlhede algorithm. Additionally we must rewrite $\i \to\i + q \bar p n\o /3a$
 in all higher order Cartan spinor invariants such as  (\ref{K2}); this will lead to changes in the coefficients of any terms containing the $n$ coordinate  in the Cartan scalar invariants (\ref{K2GHP}), (\ref{K30GHP}), (\ref{K31GHP}), (\ref{K32GHP}), (\ref{K4GHP}).

 Furthermore, we have retained in the invariants the arbitrary spin and boost parameters,  $p, q$ respectively: however, for the zeroth order in   these spacetimes the Karhede algorithm requires that $\Phi_{22}= 1$, which can be achieved with the boost choice $q=a^{1/2}$; and to complete the standard canonical  tetrad by making a real choice for $\edth \Phi_{22}$ can be achieved with the spin choice $p=1/\sqrt{2}$ .

 There is, as we noted in the Introduction,  one further difference between the GHP Cartan invariants quoted here from
GIF and the standard Cartan-Karlhede  invariants deduced from the Karlhede algorithm: the Cartan-Karlhede invariant scalars are constructed by projecting  the minimal set of covariant derivatives of the Riemannn spinor onto the frame.
Although, taking into account these various differences, we could now deduce the
Cartan-Karlhede  invariants
 by hand from the corresponding GHP  Cartan scalars just calculated, we
  shall not write  out these Cartan-Karlhede invariants explicitly here; instead we shall determine
 the   Cartan-Karlhede  invariants for this version of the spacetime by the usual CLASSI procedure, in the
next section.

\section {Karlhede classification by CLASSI of Edgar-Ludwig spacetime in intrinsic coordinates}

In this section, for the line element  (\ref {metric12}) which was obtained in \cite{edvic} we will make a coordinate change $a\to  - c $ which gives the metric in coordinates $t,n,c,b$
\begin{eqnarray}
\hbox{d} s^2 &=&\Bigl(-c\bigl(2s(t)+2cm(t)-c^2-b^2\bigr) -e(t)^2-
n^2\Bigr) \hbox{d} t^2+2c \hbox{d} t \hbox{d} n \nonumber\\
&& - 2 n \hbox{d} t \hbox{d} c + 2 e(t) \hbox{d} t  \hbox{d} b -
\hbox{d} c^2 - \hbox{d} b^2 \label{metric12c}
\end{eqnarray}
This is simply a technical change to simply the operations of the CLASSI programme in situations involving terms with square roots.

The line element  (\ref {metric12}) was obtained in \cite{edvic}
from a null tetrad  (  (67) in  \cite{edvic} ) with arbitrary spin
and boost parameters $P, Q$; more generally we now consider this
tetrad with the additional complex parameter $Z$ representing a null
rotation (and rewriting $P, Q$ as $p, q$ respectively, with $p\bar
p=1/2$),
\begin{eqnarray}\nonumber
n&=&\frac{q}{2c}\big(c^3-2c^2m(t)+cb^2-2cs(t)-e^2(t)-n^2+Z\overline{Z}\big)\hbox{d}t \\\nonumber
&&+q\hbox{d}n +\frac{q}{c}\big({\Re}\left(Z\big)-n\right)\hbox{d}c+\frac{q}{c}\big(e(t)-{\Im}\left(Z\right)\big)\hbox{d}b \\\nonumber
l&=&\frac{c}{q}\hbox{d}t, \qquad
\bar m=p\left(\overline{Z}{\hbox{d}t }+\hbox{d}c+{i\hbox{d}b}\right), \qquad
m =\bar p\left(Z{\hbox{d}t }+{\hbox{d}c}-{i\hbox{d}b}\right)\label{frame1}
\end{eqnarray}

The parameters $p, q$, ${\Re}(Z)$, $\Im(Z)$  are  used to get the Riemann tensor and its derivatives in standard form (characterised by
the functional dependence being minimised at each order). We begin with the choices $q= \sqrt{c}$ to get
$\Phi_{22^{\prime}}=1$, and
 ${\Re}(Z)=2n/3$ to get $D\Phi_{33^{\prime}}=0$\footnote{We use the abreviated notation for Cartan-Karlhede invariants used in \cite{cdI}, \cite{skea1}.}. Then $p=1/\sqrt{2}$  makes
$\Im(D\Phi_{23^{\prime}})=0$ and finally, by putting ${\Im}(Z)=e(t)$, we obtain $\Im(D^2\Phi_{34^{\prime}})=0$
 (in this way the appearance of the last
functional independent component is postponed to the third derivative). The frame is now completely specified and no isotropies remain.

The Weyl tensor and curvature scalar are zero, so the Riemann tensor is given by the $\Phi_{ab^{\prime}}$-spinor.
A classification according to the Cartan-Karlhede procedure gives at \underbar{zeroth and first order}
\begin{eqnarray}
\Phi_{22^{\prime}}=1,\qquad\qquad
D\Phi_{23^{\prime}}=-\frac{1}{\sqrt{2}c} \label{KEV0}
\end{eqnarray}
Hence the first functionally independent quantity is found in the first derivative and $c$ can be used as an
essential coordinate.

\smallskip
 \underbar{At second  order,}
\begin{eqnarray}
D^2\Phi_{24^{\prime}}&=&D^2\Phi_{33^{\prime}}=\frac{1}{c^2},\qquad \quad
D^2\Phi_{34^{\prime}}=-\frac{\sqrt{2}n}{3c^{5/2}}\nonumber\\
D^2\Phi_{44^{\prime}}&=&\frac{5}{2}-\frac{2m(t)}{c}-\frac{b^2}{2c^2}+\frac{s(t)}{c^2}+\frac{5n^2}{9c^3}
\label{KEV2}
\end{eqnarray}
Two more functionally independent components are found and all isotropy is lost; clearly $n$ can be used as a
 second essential coordinate;
we could choose $D^2\Phi_{44^{\prime}}$ as a third coordinate already at this stage,
but the choice of essential coordinates get simpler if we also use the information from $D^3\Phi$.

\smallskip
\underbar{At third  order,}
\begin{eqnarray}
D^3\Phi_{25^{\prime}}&=&D^3\Phi_{34^{\prime}}=\frac{-3}{\sqrt{2}c^3},\qquad\quad
D^3\Phi_{35^{\prime}}=D^3\Phi_{44^{\prime}}=\frac{2n}{c^{7/2}}\nonumber\\
D^3\Phi_{45^{\prime}}&=&\frac{\sqrt{2}}{c^4}\left(-\frac{37c^3}{10}+3m(t)c^2+b^2c-2s(t)c-\frac{3n^2}{2}\right)+\hbox{i} \frac{b}
{\sqrt{2}c^2}\nonumber\\
D^3\Phi_{55^{\prime}}&=&\frac{1}{c^{7/2}}\left(
  \frac{5nc^2}{3}-2c^2m'(t)-cbe(t)-2cnm(t)+cs'(t)\right.\nonumber\\
&& - \frac{5b^2n}{3}+\left.\frac{10ns(t)}{3}+\frac{5n^3}{3c}\right)
\label{KEV3}
\end{eqnarray}

From $\Im(D^3\Phi_{45^{\prime}})$ we can choose $b$ as one of the essential coordinates.

From $D^2\Phi_{44^{\prime}}$ we can then choose $m(t)$ or $s(t)$ as the last coordinate
(if at least one is non-constant). If
$m(t)=m_0$ and $s(t)=s_0$ both are constants, then  $e(t)$ (providing it is not constant)
in $D^3\Phi_{55^{\prime}}$ can be used as the fourth coordinate. On the otherhand, if  $e(t)=e_0$ also
is a constant, then  all coordinates are found in $D^2\Phi$. Note that in this case the third constant $e_0$ is found in $D^3\Phi$.

So therefore we can sum up in the same way as in the previous section,

\smallskip

$\bullet$  if at least one of the functions $m(t), s(t), e(t)$, is
not constant  then all four essential base coordinates are
obtained from Cartan-Karlhede invariants at third order, and the procedure
will therefore formally terminate at fourth order.

\smallskip

$\bullet$ when all of the functions $m(t), s(t), e(t)$ are
constants,  then only three essential base coordinates can be
obtained from Cartan-Karlhede invariants; these are obtained  at second
order, and the procedure will therefore formally terminate at
third order.

\smallskip
\underbar{At fourth  order,}
\begin{eqnarray}
D^4\Phi_{26^{\prime}}&=&D^4\Phi_{35^{\prime}}=D^4\Phi_{44^{\prime}}=\frac{6}{c^4},
\qquad \
D^4\Phi_{36^{\prime}}=D^4\Phi_{45^{\prime}}=-\frac{6\sqrt{2}n}{c^{9/2}}\nonumber\\
D^4\Phi_{46^{\prime}}&=&\frac{1}{c^4}\left(\frac{74c^2}{5}-12m(t)c-5b^2+10s(t)+\frac{10n^2}{c}\right)
-4i\frac{b}{c^3}\nonumber\\
D^4\Phi_{55^{\prime}}&=&\frac{1}{c^4}\left(\frac{265c^2}{18}-12m(t)c-5b^2+10s(t)
+\frac{10n^2}{c}\right)\nonumber\\
D^4\Phi_{56^{\prime}}&=&-\frac{161\sqrt{2}n}{18c^{5/2}}+\frac{4\sqrt{2}m'(t)}{c^{5/2}}+
\frac{5be(t)}{\sqrt{2}c^{7/2}}+\frac{8\sqrt{2}nm(t)}{c^{7/2}}-\frac{5s'(t)}{\sqrt{2}c^{7/2}}
\nonumber\\
&& + \frac{35\sqrt{2}n(b^2-2s(t))}{6c^{9/2}}-\frac{20\sqrt{2}n^3}{3c^{11/2}}+i\frac{\sqrt{2}}{c^{7/2}}\left
(\frac{e(t)c}{2}+\frac{5bn}{3}\right)\nonumber\\
D^4\Phi_{66^{\prime}}&=&30-47\frac{m(t)}{c}+\frac{1}{c^2}\left(-\frac{19}{2}b^2+18m^2(t)+29s(t)-2m''(t)\right)
\nonumber\\
&& +\frac{1}{c^3}
\left(11b^2m(t)-be'(t)-e^2(t)-22m(t)s(t)+\frac{335}{18}n^2\right. \nonumber\\
&& - 4nm'(t)+s''(t)\Bigr)
+\frac{1}{c^4}\left(\frac{5}{2}b^4-10b^2s(t)-5bne(t)-14n^2m(t)\right.
\nonumber\\
&& + 5ns'(t)+10s^2(t)\Bigr)+\frac{65n^2}{6c^5}\left( 2s(t)-b^2\right)
+\frac{145}{18}\frac{n^4}{c^6}\nonumber\\
\square(D^2\Phi_{44'})&=&{14\over c^2} 
\label{KEV4}
\end{eqnarray}

For completeness we have given all of the non-zero fourth order Cartan-Karlhede invariants
\footnote{In general the symmetrised derivatives are not sufficient for a complete classification. The additional
quantities needed are given in \cite{maccullum}. For the present case the only non-zero of these is
$\square(D^2\Phi_{44'})\equiv \Phi_{111'1';11'11'}\!^{x'}\!_a\! ^a\!_{x'}={14\over c^2}$.}. When checking for equivalence with, e.g., a metric copy as in Section 2, it is essential that all Cartan-Karlhede invariants up to fourth order are consistent. In particular, since derivatives  $s'(t), m'(t) $ appear in ${\cal R}^3$ and $e'(t)$ in ${\cal R}^4$, we need to ensure compatability of $\hbox{d}T / d\hbox{t}$ which is obtainable from each of the three functions.

\smallskip

As noted earlier, the GHP Cartan scalars used for
the classification in Section 2 are closely related to the  Cartan-Karlhede scalars determined above; by
comparing the two sets, we can see how the basic structures are the same, only the numerical
details of the coefficients differ. As pointed out before, the main difference is due to the (slightly) different
tetrads used, but also to the  difference in definitions of GHP Cartan invariant scalars and
Cartan-Karlhede scalars.

An investigation of the role of the non-redundant functions in this section
simply  duplicates the results in the previous section, and the details  give once again the same three
cases as in the previous section.

\section{{Karlhede classification by CLASSI of Edgar-Ludwig spacetime in Koutras-McIntosh  coordinates}}

Skea \cite{skea1} investigated the equivalence of the Edgar-Ludwig
spacetime
 given in \cite{edludL}, in the Koutras-McIntosh coordinates $u,w,x,y$
\begin{eqnarray}\hbox{d} s^2 = \Bigl(2f(u)x\bigl(h(u)+ g(u)y+x^2+y^2\bigr)- w^2\Bigr)\hbox{d} u^2
\nonumber\\
+ 2x \hbox{d} u \hbox{d} w - 2w\hbox{d} u \hbox{d} x - \hbox{d}
x^2- \hbox{d} y^2 \label{ELmetric}
\end{eqnarray}
or equivalently,
\begin{eqnarray} g^{ij}=\pmatrix{0 &{1/x} & 0&0\cr
{1/x} & -2f(u)\bigl(h(u)+g(u)y+x^2+y^2\bigr)/x & -{w/x} & 0\cr
0&-w/x & -1&0\cr 0&0 &0&-1} \label{ELmetric12}
\end{eqnarray}
and the Wils spacetime \cite{wils} given in the same coordinates but
with $g(u)=0=h(u)$, in order to determine whether the  Edgar-Ludwig
spacetime could actually be reduced to the Wils spacetime by a
coordinate transformation. We shall consider the more general
problem of the complete invariant classification of the Edgar-Ludwig
spacetime, in particular checking on the redundancy of the three
apparently non-redundant functions; we shall then specialise these
results to the equivalence problem of the Edgar-Ludwig and Wils
spacetimes.

The  Cartan-Karlhede scalar invariants for the Edgar-Ludwig
spacetime in Koutras-McIntosh  coordinates are given  to fourth
order in \cite{skea1}, in terms of the canonical  tetrad obtained
there, and we shall also use this tetrad and quote  directly the
Cartan-Karlhede scalar invariants in \cite{skea1}.

In \cite{skea1} it was argued that the fourth order invariant
$D^4\Phi_{56'} $ is needed to show inequivalence. However, in the
present work we find that this can rather be achieved already at
third order.

As Skea points out, Cartan-Karlhede scalar invariants of  first and second
order supply three base coordinates directly, $D\Phi_{23'}$,
$D^2\Phi_{34'} $, $D^2\Phi_{44'} $,
\begin{eqnarray} \Phi_{22'}&=&1\\
D\Phi_{23'}
&=& {-1\over \sqrt {2} x}\label{cs1}\\
D^2\Phi_{34'} &=&{xf_u-2wf\over 6f^{3/2}x^{5/2}} \label{cs3}\\
D^2\Phi_{44'} &=&-{g(u)y+h(u)\over 2x^2} +{5\over 2}-{y^2\over
2x^2}\nonumber\\
&& + {5\bigl(f(u)^2w^2-f(u)wxf_u-2x^2f_u{}^2\bigr)\over
18f(u)^3x^3}+{f_{uu}\over 2f(u)^2x} \label{cs2}
\end{eqnarray}
But instead of  $D^2\Phi_{44'} $ we shall prefer to begin with the simpler third order Cartan-Karlhede scalar
invariant,
$$\Im(D^3\Phi_{45'}) = {2y+g(u)\over 2\sqrt{2} x^2}$$

We shall now follow  the usual convention, used also by Skea, that
coordinates $x,y,w,u$ and non-redundant functions $f(u), g(u), h(u)$
in the first spacetime have direct counterparts $X,Y,W,U$ and
$F(U), G(U), H(U)$ in a second spacetime.

So if we  begin with the three simplest Cartan-Karlhede scalar invariants, we find the following equivalences
\begin{eqnarray} D\Phi_{23'}:\qquad\qquad\qquad\qquad\qquad\
x&=&X\label{ids1} \qquad\qquad\qquad\qquad\qquad\\
\Im(D^3\Phi_{45'}):\qquad\qquad\qquad\ y+g(u)/2 &=&Y + G(U)/2\qquad\qquad\qquad\qquad\qquad\label{ids2} \\
D^2\Phi_{34'}:\qquad\qquad\qquad{xf_u-2wf\over f^{3/2}}
&=&{XF_U-2WF\over F^{3/2}}
\qquad\qquad\qquad\qquad\qquad\label{ids3}\end{eqnarray} where
(\ref{ids1}) has been used to obtain  (\ref{ids2}) and
(\ref{ids3}); at this stage we have identified three essential
base coordinates, and it is obvious that they will always be
functionally independent.

Going next to the more complicated relation for $D^2\Phi_{44'}$ in
\cite{skea1}, we rearrange
\begin{eqnarray} D^2\Phi_{44'}
&=&-{g(u)y+h(u)\over 2x^2}+{5\over 2}-{y^2\over
2x^2}+{5\bigl(f(u)^2w^2-f(u)wxf_u-2x^2f_u{}^2\bigr)\over
18f(u)^3x^3} \nonumber \\ &&+{f_{uu}\over 2f(u)^2x}\nonumber\\
&=&{5\over 2}-{{\bigl(y+g(u)/2\bigr)^2 +h(u)-g(u)^2/4} \over
2x^2}\nonumber\\&&+ {5\Bigl(
\bigl(f(u)w-xf_u/2\bigr)^2-9x^2f_u^2/4\Bigr)\over 18f(u)^3x^3} +
{f_{uu}\over 2f(u)^2x}\label{cs21}
\end{eqnarray} Now, making use of  (\ref{ids1}), (\ref{ids2}) and
(\ref{ids3}), we deduce from (\ref{cs21}) that
\begin{eqnarray}F^3f^3\bigl(4h(u)-g(u)^2\bigr)+ 5x(f^3F_U^2-F^3f_u^2)+4xfF(F^2f_{uu}-f^2F_{UU})=0.
\label{ids5}\end{eqnarray}

We next consider $\Re(D^3\Phi_{45'})$ and by a similar
calculation, we find that
\begin{eqnarray}-4F^3f^3\bigl(4h(u)-g(u)^2\bigr)+ 15x(f^3F_U^2-F^3f_u^2)\nonumber\\
+ 12xfF(F^2f_{uu}-f^2F_{UU})=0. \label{ids51}\end{eqnarray}
 Together (\ref{ids5}), (\ref{ids51}) give
the key equations
\begin{eqnarray} (ff_{uu}-5f_u^2/4)f^{-3} = (FF_{UU}-5F_U^2/4)F^{-3},
\label{ids52}\end{eqnarray}
\begin{eqnarray}h(u)-g(u)^2/4 = H(U)-G(U)^2/4
\label{ids6}\end{eqnarray}
From (\ref{ids52})  we can identify a fourth
essential base coordinate --- in general, but only providing
$(ff_{uu}-5f_u^2/4 )f^{-3} \ne \hbox{constant}$; this  means that we also will need to
consider separately the case $(ff_{uu}-5f_u^2/4)f^{-3}= f_0 =
(FF_{UU}-5F_U^2/4)F^{-3}$, \ where $f_0$ is constant. Further analysis
will involve (\ref{ids6}), and again we will need to consider
separate cases $h(u)-g(u)^2/4 = h_0 = H(U)-G(U)^2/4 $ where $h_0 =$   constant, and $\ne $ constant.

The only remaining  third order Cartan-Karlhede scalar invariant
where we can get  any independent information is
$D^3\Phi_{55'}.$\footnote{Note that in the explicit expression for
$D^3\Phi_{55'}$ on page 2396 of \cite{skea1} there is a
typographical error; the first term on the right hand side should
contain the factor $(xf_u-2fw)$, and not $(x-2fw)$. $D^3\Phi_{55'}$
is given correctly in \cite{pol}.} We now carry out a similar
rearrangement as we did above for $D^2\Phi_{44'}$,
\begin{eqnarray} 2 \sqrt{2} D^3\Phi_{55'}
&=&-{g_uy+h_u\over f^{1/2}x^{5/2}} \nonumber \\&&+ {5\over
3x^{7/2}}\Bigl(\bigl(2fw-xf_u\bigr)/f^{3/2}\Bigr)
\Bigl(x^2-(y+g/2)^2-h +g^2/4\Bigr)\nonumber\\
&&+ {5\over 24
x^{9/2}}\Bigl(\bigl(2fw-xf_u\bigr)/f^{3/2}\Bigr)^3\nonumber\\
&&+ {1\over 8
x^{5/2}f^3}\Bigl(\bigl(2fw-xf_u\bigr)/f^{3/2}\Bigr) \Bigl(4f
f_{uu}-5f_u^2\Bigr)  \nonumber\\
&& + {1\over 24
x^{3/2}f^{9/2}}\Bigl(24f^2 f_{uuu}+90f_u^3- 108 f f_u f_{uu}\Bigr)
\end{eqnarray}
and, making use of  (\ref{ids1}) and  (\ref{ids3}), as well as  (\ref{ids6}), (\ref{ids52}), we deduce
\begin{eqnarray}
4x^{-1}(g_uy+h_u)f^{-1/2} - \Bigl(15 f_u^3+4f^2f_{uuu}-18 f f_uf_{uu}\Bigr)f^{-9/2}\nonumber\\
= 4x^{-1}(G_U Y+H_U)F^{-1/2} - \Bigl(15 F_U^3+4F^2F_{UUU}-18 F F_U
F_{UU}\Bigr)F^{-9/2}. \label{ids9}\end{eqnarray}
Now using
(\ref{ids2}) gives
\begin{eqnarray}
4x^{-1}y(g_uf^{-1/2} -G_UF^{-1/2})\nonumber\\
+4x^{-1}\Bigl(h_uf^{-1/2}- H_UF^{-1/2}- (gg_u f^{-1/2}- GG_UF^{-1/2})/2\Bigr) \nonumber\\ - \Bigl(15 f_u^3+4f^2f_{uuu}-18 f f_uf_{uu}\Bigr)f^{-9/2}\nonumber\\
=- \Bigl(15 F_U^3+4F^2F_{UUU}-18 F F_U F_{UU}\Bigr)F^{-9/2}
\label{ids91}\end{eqnarray}

A quick examination of the   fourth order Cartan-Karlhede
scalar invariants listed in \cite{skea1}, shows 
that most give duplications of lower order information\footnote{Note
that in the explicit expression for $D^4\Phi_{55'}$ on page 2397 of
\cite{skea1} the coefficient $265$ should be $256$.}. For the
remainder: $\Im(D^4 \Phi_{56'})$ also requires  a little manipulation to
show that no new information is available\footnote{The
Cartan-Karlhede scalar invariant $\Im(D^4\Phi_{56'})$ is missing the
additional term ${5wg\over 6x^{7/2} f^{1/2}}$}; $\Re(D^4
\Phi_{56'})$ requires a little more manipulation to show that no new
information is available\footnote{Note that in the explicit
expression for $D^4\Phi_{56'}$ on page 2397 of \cite{skea1} there is
a typographical error of the same type as in $D^3\Phi_{55'}$; the
first term on the right hand side should contain the factor
$(2fw-xf_u)$, and not $(2fw-x)$. The term in $w^3$ should have a
denominator with $x^{11/2}$ instead of $x^{1/2}$. The numerical
coefficient $63$ should instead be $36$.};  the only fourth order
Cartan-Karlhede scalar invariant which has new information is
$D^4\Phi_{66'}.$\footnote{The expression with only the term $x^4$ in
denominator should instead have $4 x^4$.} We now carry out a
similar rearrangement as we did above for $D^2\Phi_{44'}$ and
$D^3\Phi_{55'}$, and making use of all earlier results,  obtain
\begin{eqnarray} f_{uuuu}/(4f^3)-135f_u^4/(128f^6)-45f_u^3w/(16xf^5)\nonumber \\-9f_{uu}f_u^2/(4f^5)+27f_{uu}f_uw/(8f^4x)-11f_{uuu}f_u/(4f^4)\nonumber\\
-3f_{uuu}w/(4xf^3)+(g_uy+h_u)f_u/(8xf)-(g_{uu}y+h_{uu})/(4fx)\nonumber\\
=F_{UUUU}/(4F^3)-135F_U^4/(128F^6)-45F_U^3W/(16XF^5)\nonumber \\-9F_{UU}F_U^2/(4F^5)+27F_{UU}F_UW/(8F^4X)-11F_{UUU}F_U/(4F^4)\nonumber\\-3F_{UUU}W/(4XF^3)
+(G_UY+H_U)F_U/(8XF)-(G_{UU}Y+H_{UU})/(4XF)\label{fourth}
\end{eqnarray}

It is clear that a complete invariant classification of this
spacetime in these coordinates will involve a generic case, and a
number of special cases.

So we now look at each of these various
cases individually, and in full detail.

\smallskip

{\bf Case (A)} \ $(ff_{uu}-5f_u^2/4 )f^{-3} \ne $ constant.

For this case, (\ref{ids52}) gives a fourth   essential base coordinate $U$,
which is a function of $u$ alone.
We have found four essential base coordinates in ${\cal R}^3$; so the algorithm will formally
terminate at  ${\cal R}^4$.
\smallskip

{\bf Case (B)}\  $(ff_{uu}-5f_u^2/4)f^{-3} = f_0 $, {constant}; \
$h(u)-g(u)^2/4 \ne$ constant.

For this case,  from (\ref{ids6}) we get a fourth essential base coordinate $U$,
which is a function of $u$ alone.
Again, we have found four essential base coordinates in ${\cal R}^3$; so the algorithm will formally
terminate at  ${\cal R}^4$.

\smallskip

{\bf Case (C)}  $(ff_{uu}-5f_u^2/4)f^{-3} = f_0 $, {constant}; \
$h(u)-g(u)^2/4 = h_0$ constant.

There are only three
essential base coordinates at this stage; but when these constants
are substituted into (\ref{ids91}) we obtain
 \begin{eqnarray} F^{-1/2}G_U = f^{-1/2}g_u\label{fg=FG}
 \end{eqnarray}
which supplies a fourth essential  base coordinate --- but only providing that
$f^{-1/2}g_u \ne $ constant.

Now we have to split into two subcases:

\smallskip
{\bf (i)} If  $f^{-1/2}g_u \ne $ constant,  then from (\ref{fg=FG})
we have a fourth essential  coordinate, and can conclude $U=U(u)$.
Again, we have found four essential base coordinates in ${\cal R}^3$; so the algorithm will formally terminate at  ${\cal R}^4$.

\smallskip
{\bf (ii)} If  $f^{-1/2}g_u = g_0$, constant, then  we have  no fourth essential  coordinate.
We note that no expressions
containing the $u$ coordinate occur in the Cartan-Karlhede scalar
invariants; hence  there is no fourth base coordinate  for this
special case, in ${\cal R}^3$. In fact, for this special case,
it can be seen  that the three essential base coordinates are all
given in ${\cal R}^2$  in (\ref{cs1}), (\ref{cs3}), (\ref{cs2}),
and since there is no further information about essential
coordinates supplied at third order, the procedure terminates at  ${\cal R}^3$ for this special case.

\subsection{The three apparently non-redundant  functions.}

Investigations of the nature of the three apparently non-redundant
functions in this coordinate system are more complicated and
involved than in the intrinsic coordinates earlier. We will
consider each case separately in detail:

\smallskip

{\bf Case (A)}

 The identification of a fourth essential base coordinate $U= U(u)$ enables us to  separate
out those terms with different coordinates in (\ref{ids91}),  to
get
\begin{eqnarray}\Bigl(15 f_u^3+4f^2f_{uuu}-18 f f_uf_{uu}\Bigr)f^{-9/2}
\nonumber \\= \Bigl(15 F_U^3 +4F^2F_{UUU} -18 F F_U F_{UU}\Bigr)F^{-9/2}
\label{ids10}\end{eqnarray}
\begin{eqnarray}
f^{-1/2}g_u = F^{-1/2}G_U,   \label{ids101} \end{eqnarray}
\begin{eqnarray}
f^{-1/2}h_u =  F^{-1/2} H_U +(g-G)G_UF^{-1/2}/2. \label{ids102} \end{eqnarray}

Moreover, when we differentiate (\ref{ids52})
  with respect to the coordinate $U(u)$ we obtain
\begin{eqnarray}
\Bigl(15 f_u^3+4f^2f_{uuu}-18 f f_uf_{uu}\Bigr)f^{-4}{\hbox{d}
u\over \hbox{d} U} \nonumber\\= \Bigl(15 F_U^3+4F^2F_{UUU}-18 F
F_U F_{UU}\Bigr)F^{-4}. \label{ids11}\end{eqnarray} Comparing
(\ref{ids11}) with (\ref{ids10}) gives
\begin{eqnarray}
{\hbox{d} u\over \hbox{d} U}= {F^{1/2}\over f^{1/2}} .
\label{ids12}\end{eqnarray}
The equation (\ref{ids52}) can now be
rearranged, using (\ref{ids12}), into a second order differential
equation for $F(U)$ in terms of $f(u(U))$, with the solution
\begin{eqnarray} F(U)= f(u(U))/ (c_0+c_1 U )^4, \qquad c_1, c_0 \ \hbox{ are constants}\label{F=f}\end{eqnarray}
From (\ref{F=f}) it follows that
\begin{eqnarray}
{\hbox{d} u\over \hbox{d} U}= { (c_0+c_1 U )^{-2}},
\label{ids121}\end{eqnarray}
so that, from (\ref{ids121}),
\begin{eqnarray} u(U)= c_2 -{1\over c_1(c_0+c_1 U )}, \   \hbox{for}  \  c_1\ne 0, \quad \hbox{and} \quad  u(U) = c_2 +{U\over c_0^2 }, \  \hbox{for} \   c_1= 0 \label{U=u}\end{eqnarray}
for $c_2$ constant; and hence
\begin{eqnarray}F(U)&=&{1\over (c_0+c_1 U )^{4}}f\bigl(c_2 -{1\over c_1(c_0+c_1 U )}\bigr), \
\hbox{for}  \  c_1\ne 0,\nonumber\\ F(U)& = & {1\over
c_0^4}f\bigl(c_2 +{U\over c_0^2 }\bigr), \  \hbox{for} \   c_1= 0 .
\label{ff=FF}
\end{eqnarray}

In addition, using (\ref{ids12}) in (\ref{ids101}), we get
\begin{eqnarray}
G_U=g_U,  \quad \hbox{and hence}, \quad  G(U)=g(u(U)) + c_3, \ \hbox{where $c_3$ is constant}, \label{ids1011} \end{eqnarray}
and where we can replace $u(U)$ with the two possibilities in (\ref{U=u}).

There remains still some more information in (\ref{ids102}): combining with (\ref{ids12})
\begin{eqnarray}
 H_U=h_U - (g-G) G_U/2,  \label{ids1012} \end{eqnarray}
but it is easy to see that  (\ref{ids1012}) is just the $U$
derivative of (\ref{ids6}); therefore, from (\ref{ids6}),
\begin{eqnarray} H(U)=  h(u(U))+c_3^2/4+c_3g(u(U))/2\label{hh=HH}\end{eqnarray}
 where we can replace $u(U)$ with the two possibilities in (\ref{U=u}).

Clearly since $f(u), g(u), h(u)$ and $F(U), G(U), H(U)$ are  completely arbitrary functions,
then (\ref{ff=FF}), (\ref{ids1011}),   (\ref{hh=HH}) respectively    can always be satisfied  for a given choice of $f(u), g(u), h(u)$ or $F(U), G(U), H(U)$. According to the theory, incompatability might appear at next order; but a direct check confirms that the fourth order conditions are identically satisfied, and hence we have  equivalence.

\smallskip

{\bf Case (B)}

In this case $F(U)$ can be obtained directly, since from
(\ref{ids52}) we obtain also $(FF_{UU}-5F_U^2/4)F^{-3} = f_0 $.
These can be integrated to give \cite{skea1}
 \begin{eqnarray}
 f(u) = {c_0^2\over (c_0^2(u-u_0)^2-f_0)^2}\ \hbox{for} \  c_0, u_0 \ \hbox{constants}, \label{intf}\\
  F(U) = {C_0^2\over (C_0^2(U-U_0)^2-F_0)^2}\ \hbox{for} \  C_0, U_0 \ \hbox{constants} . \label{intF}
\end{eqnarray}

As in the previous case, since $U=U(u)$ we can separate  (\ref{ids91}) to get
(\ref{ids101}) and  (\ref{ids102}).

 ( Condition { (\ref{ids10}) is trivially satisfied, since, in this case, we have the trivial identity,
$$15 f_u^3
+4f^2f_{uuu} -18 f f_u f_{uu}= 0 =15 F_U^3 +4F^2F_{UUU} -18 F F_U
F_{UU} . \  \ )$$
The derivative with respect to the fourth coordinate $U = U(u)$ of
(\ref{ids6}) is
\begin{eqnarray}(4H_U-2GG_U)=(4h_u-2gg_u){\hbox{d} u\over \hbox{d} U}
\label{dHG}\end{eqnarray} and comparing with (\ref{ids102}) gives again (\ref{ids12});
 and so,  from  (\ref{intf}),  (\ref{intF}) we can find the relationship between the fourth pair of coordinates from
 \begin{eqnarray} c_0\int { \hbox{d} u \over c_0^2(u-u_0)^2-f_0} = C_0\int { \hbox{d} U\over C_0^2(U-U_0)^2-f_0}   \ ,
 \label{U2=u2}
 \end{eqnarray}
where there will be different explicit integrals depending on the nature of $f_0$.
 Also from (\ref{ids12}), together with  (\ref{ids101}), we obtain again (\ref{ids1011}),
and therefore, as in the generic Case {\bf (A)}, we obtain $G(U), H(U)$ from (\ref{ids1011}) and (\ref{hh=HH}) respectively, where we can replace $U(u)$ from the results in (\ref{U2=u2}).

Clearly since $f(u), g(u)$ and $F(U), G(U)$ are  completely arbitrary functions,
then (\ref{ff=FF}), (\ref{ids1011}),    respectively    can always be satisfied  for a given choice
of $f(u), g(u)$ or $F(U), G(U)$; furthermore, for the special case of
$f(u)$ defined by (\ref{intf}), the  corresponding $F(U)$ is
defined by (\ref{intF}).  Once again the fourth order conditions are identically satisfied and we have equivalence.

\smallskip

{\bf Case (C)}

In this case --- as in Case {\bf (B)} --- $F(U)$ and $f(u)$ can be obtained directly, and are given
by (\ref{intf}) and (\ref{intF}). We also obtain directly
 \begin{eqnarray}
  h(u)- g(u)^2/4 =h_0 =  H(U)- G(U)^2/4\label{inth}
\end{eqnarray}
for $h_0$ an arbitrary constant.

\smallskip
{\bf (i)} $f^{-1/2}g_u \ne $ constant

For this subcase, we cannot get any more direct information  in ${\cal R}^3$; so we substitute all the above identities for this subcase in the fourth order invariant (\ref{fourth}),
and after a long simplification obtain
\begin{eqnarray}F^{-1}G_{UU} - F^{-2}F_UG_U/2 = f^{-1}g_{uu} -
 f^{-2}f_ug_u/2 .\label{fourths}
 \end{eqnarray}
Moreover, when we differentiate (\ref
{fg=FG}) with respect to the fourth coordinate $U=U(u)$, we get
 \begin{eqnarray}F^{-1/2}G_{UU} - F^{-3/2}F_UG_U/2 =  \bigl(f^{-1/2}g_{uu} -
 f^{-3/2}f_ug_u/2\bigr){\hbox{d} u\over \hbox{d} U} ,\label{fg1=FG1}
 \end{eqnarray}
  and so we once again obtain (\ref{ids12}), and hence from (\ref{intf}) and   (\ref{intF}) we can obtain again
the relationship (\ref{U2=u2}) between the fourth coordinates. In addition, (\ref{ids12}), together with  (\ref{ids101}), leads again to (\ref{ids1011}), where we can replace $U(u)$ from the results in (\ref{U2=u2}).

Clearly since $g(u)$ and $G(U)$ are  completely arbitrary functions,
then  (\ref{ids1011})   can always be satisfied  for a given choice of $g(u)$ or $G(U)$; furthermore,
for the special case of $f(u),  h(u)$ defined by (\ref{intf}),
 (\ref{inth}) the corresponding $F(U),  H(U)$ are given by (\ref{intF}),
(\ref{inth}).
 In this particular case, we do actually get new information from the fourth order Cartan-Karlhede invariants;
  one non-trivial {\it fourth} order condition (\ref{fourth})  was needed to complete the
   invariant classification via (\ref{fourths}) in order to complete the information about the one remaining non-redundant
   function, $G(U)$, but the remaining fourth order invariants are identically satisfied. Hence we have  equivalence.

\smallskip

{\bf (ii)} $f^{-1/2}g_u = g_0, $ constant

For this subcase, since the functions $f(u), F(U)$ are given
by (\ref{intf}) and (\ref{intF}),  we can use these to get immediately
 \begin{eqnarray}
 g(u)= c_0g_0\int { \hbox{d} u \over c_0^2(u-u_0)^2-f_0} \  , \
G(U)= C_0 g_0 \int { \hbox{d} U \over C_0^2(U-U_0)^2-f_0} \label{intG}
 \end{eqnarray}
where there will be different explicit integrals depending on the
nature of $f_0$; these are quoted explicitly in \cite{skea1}. The
functions $h(u)$ and $H(U)$ follow from (\ref{inth}).

As noted above, for this special subcase,  the three essential base
coordinates are given in ${\cal R}^2$, and the procedure formally
terminates in ${\cal R}^3$. So all that remains is to relate the fourth pair of (non-essential) coordinates. Note that this cannot be found from ${\cal R}^3$, so  we compare the line element (\ref{ELmetric}) with its direct copy in $U,W,X,Y$ coordinates; when we make the substitutions which we have just identified,
\begin{eqnarray}x= X, \qquad y= Y +c_3/2 , \qquad w = {f^{1/2}\over 2}\Bigl( {2W\over F^{1/2}}+x({f_u\over f^{3/2}} -{F_U\over F^{3/2}})\Bigr)\label{coordtr}
\end{eqnarray}
where $c_3=G(U)-g(u) =  \hbox{constant}$.  It follows, after considerable simplification,   that
\begin{eqnarray}{\hbox{d}U\over \hbox{d}u} = {f^{1/2}\over F^{1/2}}
\label{dU=du1}
\end{eqnarray}
as in the other cases;  since $F(U), f(u)$ are known in this special
case, this  yields again (\ref{U2=u2}).

For this special subcase,    the
functions $f(u), g(u), h(u)$ defined by (\ref{intf}),
(\ref{intG}), (\ref{inth}) have the corresponding functions $F(U), G(U), H(U)$  given by (\ref{intF}), (\ref{intG}),
(\ref{inth}).  The three essential coordinates have been identified
in ${\cal R}^2$, and the functional relationships are given in  ${\cal R}^3$ where there is complete compatability.
Hence there is equivalence also for this special subclass.

\smallskip
So overall there is  equivalence in all three cases, with the
apparently non-redundant functions (including the special cases when they are non-redundant constants)
genuinely non-redundant; furthermore, there was only one subcase, Case
{\bf (C(i))} which required explicit use of fourth order Cartan
scalar invariants.

\medskip

The coordinate freedom for the metric in these coordinates is, in general, given by (\ref{coordtr})
with the fourth coordinate pair having different transformations depending on the different cases:
(\ref {U=u}) for Case {\bf (A)}, and (\ref{U2=u2}) for Cases {\bf (B)}, {\bf (C)},
and hence also different relationships, for these different cases,  between the respective
sets of non-redundant functions.

There was not the direct trivial identifications between the coordinates and the  sets of non-redundant
functions such as we obtained in the intrinsic coordinate version in the previous sections; this is because
the Koutras-McIntosh  coordinates and tetrad have  less of an  intrinsic character.

\subsection{ Equivalence problem for the Edgar-Ludwig and Wils
spacetimes}

In order to specialise the work in this section to the equivalence problem for the
Edgar-Ludwig and Wils spacetimes, we put $H(U)=0=G(U)$ for the
latter.  It is obvious from the above calculations, that inequivalence can be deduced at an  early stage in the
above calculations --- in fact from equation (\ref{ids6}); this is
in ${\cal R}^3$. (Although a fourth
order invariant gave new information in the  invariant classification
of the Edgar-Ludwig spacetime --- for the special subclass Case
{\bf (C(i))} --- this subclass does not intersect with the Wils
metric.)

Of course the Edgar-Ludwig spacetime with $h(u)=0=g(u)$ coincides
with the Wils metric, but, in view of the fact that there was a non-trivial
identification between the respective sets of non-redundant functions, we would expect a
more general subclass of the Edgar-Ludwig spacetime with the property of the Wils spacetime.

As well as equation  (\ref{ids6}), which we have just noted, which imposes
\begin{eqnarray} h(u)=g(u)^2/4
\end{eqnarray}
we have to consider equation  (\ref{ids52}),
\begin{eqnarray} (ff_{uu}-5f_u^2/4)f^{-3} = (FF_{UU}-5F_U^2/4)F^{-3}.
\label{ids521}
\end{eqnarray}
These two conditions imply that the subclass must be in  Case {\bf (A)} and Case {\bf (C(ii))}.

 From (\ref{ids101}) (Case {\bf (A)}) and (\ref{fg=FG}) (Case {\bf (C(ii)}) it immediately follows that
\begin{eqnarray} g_u =0, \quad \hbox{hence}\qquad g(u) \ \hbox{and}\  h(u) \  \hbox {are constants} .
\label{ids999}\end{eqnarray}
When these results are substituted,  the remaining
 second and third order Cartan-Karlhede invariants are identically satisfied.
 Further substitutions show that all fourth order Cartan-Karlhede invariants are identically satisfied.
This confirms Skea's result \cite{skea1} that the Edgar-Ludwig
spacetime only reduces to the Wils metric when $g(u)$ and $h(u)$ are
constant functions; but it has also been shown here that this
conclusion can be obtained by the {\it third order} of
Cartan-Karlhede invariants.

\smallskip

\subsection{Comparison with Skea's approach}

Finally we will compare the details of the above analysis with the
arguments in \cite{skea1}. Early in the analysis, it was argued that
itwas necessary to use the fourth order Cartan-Karlhede scalar invariant
$\Im(D^4\Phi_{56'})$ for an invariant analysis of the Edgar-Ludwig
spacetime, and to solve the equivalence problem between this
spacetime and the Wils spacetime; however, we have shown that this
invariant is not independent of lower order invariants,
 and gives no new information.

In his analysis of equation (\ref{ids52}), Skea concentrates on the generic case and
 does not consider explicitly  the possibility of $(ff_{uu}-5f_u^2/4)f^{-3} = f_o$,  constant,
    and other subcases defined by constant functions --- at that stage. However in order to consider the Killing vector case
(our Case {\bf (C(ii))}), he is led to the equivalent equation $15
f_u^3 +4f^2f_{uuu} -18 f f_u f_{uu}= 0 ,$ and to other equations
defining subcases. Subsequently he  presents a table identifying the
three cases corresponding to our Cases {\bf (A), (B), (C(i))}.

\section{Equivalence problem between metric in intrinsic coordinates and
Koutras-McIntosh coordinates}

We shall now solve the equivalence problem for the Edgar-Ludwig spacetime given in the two different coordinate systems, using the respective invariant classifications   in Sections 3 and 4. The complete
result will give the explicit coordinate transformations between
the two systems.

The metrics (\ref{metric12c}) and (\ref{ELmetric}) should be equivalent and we now proceed to show this by comparing their
invariant classifications, i.e., by solving the set of equations
\begin{eqnarray}
D^n\Phi_{ij^{\prime}}=D^n\tilde\Phi_{ij^{\prime}}
\end{eqnarray}
where $\Phi_{ij^{\prime}}$ and $\tilde\Phi_{ij^{\prime}}$ refer to the two different metrics respectively.

\smallskip
\underbar{At first order,}
from $D\Phi_{23^{\prime}}$ and $ D\tilde \Phi_{23^{\prime}}$ it follows that
\begin{eqnarray}
c=x.
\label{eq1}
\end{eqnarray}
\underbar{At second order,}
from $D^2\Phi_{34^{\prime}}$ and $ D^2\tilde\Phi_{34^{\prime}}$ it follows that
\begin{eqnarray}
n=\frac{2fw-xf_{u}}{(2f)^{3/2}}.
\label{eq2}
\end{eqnarray}
One also obtains  from $D^2\Phi_{44^{\prime}}$ and $ D^2\tilde\Phi_{44^{\prime}}$,
\begin{eqnarray}
\frac{4ff_{uu}-5f_{u}^2}{4f^3}x+b^2-\left(y+\frac{g}{2}\right)^2+\frac{g^2}{4}-h+4mx-2s=0
\label{eq21}
\end{eqnarray}
\smallskip
\underbar{At third order,} from $\Im(D^3\Phi_{45^{\prime}})$ and $\Im D^3(\tilde\Phi_{45^{\prime}})$ it follows that
\begin{eqnarray}
b=y+\frac{g}{2} .
\label{eq3}
\end{eqnarray}
At this stage we have related three pairs of essential coordinates.

From $\Re(D^3\Phi_{45^{\prime}})$ together with $D^2\Phi_{44^{\prime}}$ and their counterparts, one
then gets
\begin{eqnarray}
m(t)=\frac{5f_{u}^2-4ff_{uu}}{16f^3}, \qquad\qquad  s(t)=-{(h - g^2/4)\over 2}
\label{ids55}
\end{eqnarray}

\noindent
Making use of the above results, from
$D^3\Phi_{55^{\prime}}$  and $D^3\Phi_{55^{\prime}}$ and their counterparts, we obtain
\begin{eqnarray}
8\sqrt{2}\Bigl(-2x^2m'(t)-x\bigl(y+g/2 \bigr)e(t) + xs'(t)\Bigr)
\nonumber \\ =  -4(g_uy+h_u)xf^{-1/2} + x^2\Bigl(15 f_u^3+4f^2f_{uuu}-18 f f_uf_{uu}\Bigr)f^{-9/2}\label{eq32}
\end{eqnarray}

A full analysis of these identities will involve separate consideration of constant and non-constant functions,
as in the previous sections.
So  looking at the three cases from Section 4, in turn:

\smallskip
{\bf Case (A)} ${(ff_{uu}-5f_{u}^2/4)}f^{-3} \ne f_0$, constant.

For this case, from (\ref{ids55}) we deduce a fourth essential coordinate $t$,
which is a function of $u$ alone, and from (\ref{eq32}), by separation it follows that
\begin{eqnarray}
 m'(t)=-\Bigl(15 f_u^3+4f^2f_{uuu}-18 f f_uf_{uu}\Bigr)f^{-9/2}/16
\label{epdm}\\
s'(t)=  f^{-1/2}(2g g_u-4h_u ),
\label{epds}\\
e(t)=g_u /2\sqrt{2}f^{1/2}.
\label{K323} \end{eqnarray}

Furthermore, by differentiating (\ref{ids55}) with respect to $t$, and comparing with (\ref{epdm}), we obtain
\begin{eqnarray}
{\hbox{d}u\over \hbox{d}t} = {1\over \sqrt{2}f^{1/2}}, \qquad \hbox{and hence}, \quad  t = \sqrt{2} \int f(u)^{1/2} \hbox{d}u ,\label{ids81}
\end{eqnarray}
from which we can  invert and get the explicit relationship $u=u(t)$ for any function $f(u)$.  The functions $m(t), s(t), e(t)$ are
then determined by making the substitution $u=u(t)$  into the equations (\ref{ids55}), (\ref{K323}). As noted above,
formally we need to go one order further, but it is easy to confirm that all the
 Cartan-Karlhede invariants   are identically satisfied at fourth order.

\smallskip
{\bf Case (B)} ${(5f_{u}^2-4ff_{uu})}f^{-3} = f_0$, constant; \
$h(u)-g(u)^2/4 \ne$ constant.

This, as we saw in Section 4,  corresponds to
 \begin{eqnarray}
 f(u) = { c_o^2 \over (c_0^2(u-u_0)^2-f_0)^{2}} \ \hbox{for} \  c_0, u_0 \ \hbox{constants} .\label{intf1}
\end{eqnarray}
For this case,  from (\ref{ids55}) we get a fourth essential coordinate $t$ , and we can conclude that $t$ is a function of $u$ alone, and also that $m(t)=f_0/16$; from (\ref{eq32}), by separation, there follows again (\ref{epds}) and (\ref{K323}).

Furthermore, by differentiating (\ref{ids55}) with respect to $t$, and comparing with (\ref{epds}), we obtain once again (\ref{ids81}),
from which we can get
\begin{eqnarray}
t =  \sqrt{2} c_0 \int { \hbox{d}u \over c_0^2(u-u_0)^2-f_0} ,\label{ids911f}
\end{eqnarray}
where there will be different explicit integrals depending on the nature of $f_0$; by inversion we obtain $u(t)$.

  The functions $s(t), e(t)$ are
then determined by making the substitution $u=u(t)$  into the equations (\ref{ids55}), (\ref{K323}). As noted above,
formally we need to go one order further, but it is easy to confirm that all the
 Cartan-Karlhede invariants  are identically satisfied at fourth order.

\smallskip
Hence Cases {\bf (A), (B)} in Section 4 together correspond directly to Case {\bf (a)}
 in Sections 2,3.

\smallskip
{\bf Case (C)}  $(ff_{uu}-5f_u^2/4)f^{-3} = f_0 $, {constant}; \
$h(u)-g(u)^2/4 = h_0$, constant.

These conditions, as we saw in Section 4,  correspond to (\ref{intf1}) and
 \begin{eqnarray}
 h(u) = h_0 + g(u)^2/4 \ .\label{inth1}
\end{eqnarray}

For this case,  from (\ref{ids55}), it follows that $m(t)=f_0/16,$ and $s(t)=h_0/2$, and hence from   (\ref{eq32}) that
\begin{eqnarray}
e(t)=g_u /2\sqrt{2}f^{1/2}.
\label{K3241}
\end{eqnarray}

\smallskip

{\bf (i)} $f^{-1/2}g_u \ne$ constant.

From (\ref{K3241}) we get a fourth essential coordinate $t$, and we can conclude that $t$ is a function of $u$ alone. There is no more information in ${\cal R}^3$.

At \underbar{fourth order}, when we equate the invariants $D^4\Phi_{66'}$ and $D^4\tilde\Phi_{66'}$, and obtain the major simplification by substituting in all the information from this subcase, we obtain
\begin{eqnarray}
e'(t)=(2f^{-1}g_{uu}-f^{-2}f_ug_u)/8 .
\label{K3242} \end{eqnarray}
Then, by  differentiating (\ref{K3241}) with respect to $t(u)$, and comparing with (\ref{K3242}), we obtain again (\ref{ids81})
from which we can get again (\ref{ids911f}). The function $e(t)$ is then determined by making the substitution $u=u(t)$  into the equation  (\ref{K3241}).
Hence we need to go to fourth order of invariants to complete the equivalence problem for this case; furthermore, it is easy to confirm that there is no more information in ${\cal R}^4$.

Hence Case {\bf (C(i))} in Section 4 corresponds to Case {\bf (b)}
 in Sections 3,4.

\smallskip
{\bf (ii)} $f^{-1/2}g_u =g_0,$ constant. 

This additional  condition, as we saw in Section 4,  corresponds to
 \begin{eqnarray}
 g(u)=  c_0 g_0\int {\hbox{d} u \over c_0^2(u-u_0)^2-f_0} \ . \label{intg1}
\end{eqnarray}
In addition, it follows from (\ref{K3241}) that $e(t)= g_0/2\sqrt{2}$, and so all three functions $m(t), s(t), e(t)$ are constant in this case.

Hence Case {\bf (C(ii))} in Section 4 corresponds to Case {\bf (c)}
 in Sections 3,4.
(As noted in the earlier sections, the three essential base coordinates are
given in ${\cal R}^2$
and since there is no further information about essential
coordinates supplied at third order, the procedure formally terminates in ${\cal R}^3$ for this special case.)

In this case it  remains to identify the non-essential coordinate, $ t$. To do this we compare the line elements (\ref{metric12c}) and (\ref{ELmetric12}),
by making the substitutions which we have just identified,
\begin{eqnarray}c=x,\qquad b= y +g/2, \qquad  n = {2fw -xf_u\over (2 f)^{3/2}}\label{coordtr1}
\end{eqnarray}
where $f(u)$ is given by (\ref{intf1}) and $g(u)$ by (\ref{intg1}).  After  considerable simplification, it emerges that
\begin{eqnarray}{\hbox{d}t\over \hbox{d}u} = \sqrt{2} {f^{1/2}}
\label{dt=du}
\end{eqnarray}
as in the other cases;  since $f(u)$ is known in this special case, this  yields again (\ref{ids911f}).

 \medskip
So overall there is  equivalence in all cases, with the
apparently non-redundant functions (including the special cases when they are non-redundant constants)
genuinely non-redundant.

\medskip

The coordinate freedom for the metric in these coordinates is, in general, given by (\ref{coordtr1})
with the fourth coordinate pair having different transformations depending on the different cases:
(\ref{ids81}) for Case {\bf (A)}, and (\ref{ids911f}) for Cases {\bf (B)}, {\bf (C)},
and hence also different relationships, for these different cases,  between the respective
sets of non-redundant functions.

\section{Symmetries}

Barnes has, by a long  direct calculation \cite{barnes}, integrated
the conformal Killing equations for the Edgar-Ludwig metric. From
this, he has identified the special cases for a Killing vector, and
a homothetic Killing vector. He also identified the very special
case involving a two dimensional homothety group;   this case had
been overlooked in a  previous investigation by Edgar and Ludwig  in
\cite{edlud2}. The coordinates which Barnes uses are very close to
the intrinsic coordinates used in Section 2 and 3, and this has
meant that his final results are in a comparatively simple form. To
carry out an analogous direct integration using the Koutras-McIntosh
coordinate version of the metric would be considerably longer and
more complicated, and  the presentation of the final results would
also be in a more complicated form.

However, when we take into account the nature of the coordinates and
tetrad in the {\it intrinsic coordinate version} of the metric
(\ref{metric12}), it is not necessary to integrate the Killing
equations directly, in order to obtain a symmetry analysis. We
illustrate, in the next two subsections, how simple such a symmetry
analysis for Killing and homothetic Killing vectors can be in
intrinsic coordinates, and compare with the much longer and more complicated Killing vector analysis
for the Koutras-McIntosh coordinate version  in the final
subsection.

\subsection{Killing vectors in intrinsic coordinate version}

An efficient way to investigate Killing vectors and homothetic
Killing vectors in the GHP formalism has been developed in
\cite{edlud2}. An {\it intrinsic  GHP tetrad}  is a tetrad where the
vector directions are fixed  by the Riemann tensor and its
derivatives; the {\it intrinsic GHP scalars} are defined, with
respect to an intrinsic GHP tetrad, to be the well-behaved GHP spin
coefficients, all the Riemann tensor tetrad components, all the GHP
derivatives of these spin coefficients and of the Riemann tensor
tetrad components, together with properly weighted functional
combinations of all of these. It has been shown in \cite{edlud2},
that all intrinsic  GHP scalars of zero spin and boost weight
$\eta$, will be Lie derived by any Killing vector $\bfxi$ present in
the spacetime under consideration, i.e., $\xi^i \eta_{,i} = 0 $.

In the version of the metric (\ref{metric1}) found using the GIF
and analysed in Section 2, the canonical
tetrad used  is obviously an intrinsic GHP tetrad; moreover,  the GHP Cartan
invariants are clearly intrinsic GHP scalars, and since they  are directly identified with GHP Cartan
invariants, so also are
the essential base coordinates $a,b,n$.  Since in addition $a,b,n$ have zero spin and boost weight,  they must be  Lie derived by any Killing vector $\bfxi$ present, i.e., $\xi^i
a_{,i} = 0, \ \xi^i
b_{,i} = 0, \ \xi^i
n_{,i} = 0$.

When there is at least one  function of
$t$ which is not constant, then by a simple coordinate transformation, e.g. $s(t)\to t$,  $t$ transforms into a fourth base coordinate
which can also be directly identified as a GHP Cartan scalar invariant, and hence $\xi^i
t_{,i} = 0$.  However,
since this implies that all {\it four} coordinates would have to be Lie derived by any
Killing vector, no Killing vector $\bfxi$ can exist in these circumstances, when at least one of the functions is non-constant.

On the other hand,
when all three  functions are constant, (respectively $m_0, s_0, e_0$) then $t$ is a
cyclic coordinate not connected to a GHP Cartan invariant,  and hence $\xi^i
t_{,i} \ne 0$, and so there is one Killing vector, which can be scaled to
 $\bfxi = {\partial\over \partial t}$, for  the  special case of (\ref{metric12})  given by
\begin{eqnarray}
\hbox{d} s^2&=&\Bigl(a\bigl(2s_0-2am_0-a^2-b^2\bigr) -e_0^2-
n^2\Bigr) \hbox{d} t^2-2a \hbox{d} t \hbox{d} n + 2 n \hbox{d} t
\hbox{d} a \nonumber\\
&& + 2 e_0 \hbox{d} t  \hbox{d} b - \hbox{d} a^2 - \hbox{d} b^2 \
.\label{metric12k}
\end{eqnarray}

 A fuller discussion of these Killing vector arguments in the GHP formalism, in these coordinates,
 is given in \cite{edlud3}.

\subsection{Homothetic Killing vector in intrinsic
coordinate version}

Koutras and Skea \cite{ks} have given an algorithm to determine
whether a spacetime admits a homothetic vector\footnote{Exceptional
spacetimes, to which this algorithm is not applicable, are
generalised plane waves and homogeneous spacetimes.}. This algorithm
was designed to exploit an invariant classification using the
Karlhede algorithm, but it is easy to see that it can be used for
any invariant classification, as given below:

\smallskip

(1) Use an existing classification algorithm to provide an invariant classification of the spacetime, ${\cal R}^n$.

(2) Choose two nonzero elements of ${\cal R}^n$ which are {\it not} both invariant under boosts, and choose a boost which sets the ratio of these elements, raised to their inverse conformal weight, constant.

(3) In this boosted basis, form the sets of ratios of all nonzero members of ${\cal R}^n$ raised to their inverse weights.  Call this set ${\cal S}^n$.

(4) Calculate the number of functionally independent functions of the coordinates in ${\cal S}^n$.

(5) If this number is one less than the number of functions of the coordinates in ${\cal R}^n$,
the spacetime is homothetic. Otherwise the space-time is not homothetic.

\smallskip

In fact, it is very simple to apply this algorithm in the  GHP
formalism since it is invariant under spin and boost
transformations. Therefore, to illustrate this procedure in these
formalisms,
 we shall use the algorithm on the invariant classification in Section 2.

In this particular case, for simplicity we specialise the spin
parameter $p=1/\sqrt{2}$. Since  $\Phi$  has
conformal weight $-2$ and boost weight $2$, whereas $\edth
\Phi$ has conformal weight $-3$
and boost weight $2$, we note that
$$(\Phi)^{-1/2} = (q^2/a)^{-1/2}, \quad \hbox{and} \quad (\edth
\Phi)^{-1/3} = 2^{1/6}( q^2/a^2)^{-1/3}
$$
and so to fulfill step (2) of the algorithm we  specialise the boost
$$q=a^{-1/2} .$$

 The remaining non-trivial  GHP Cartan scalars
when specialised with this boost are as follows:

\underbar {First order with conformal weight $-3$.}
\begin{eqnarray}\thorn'\Phi= -a^{-3} (n a^{-3/2}).
\end{eqnarray}
\underbar {Second order with conformal weight $-4$.}
\begin{eqnarray}\edth \edth \Phi&=&a^{-4},\quad (\edth'\edth \Phi)={1\over 2}a^{-4},\quad (\edth \thorn'
\Phi)  = -a^{-4}{3\over \sqrt{2}}   (n a^{-3/2}),\\
 \thorn' \thorn' \Phi
&=& a^{-4} \Bigl({1\over 2} +3 (n a^{-3/2})^2 + {1\over 2}(b
a^{-1})^2 -\bigl(s(t) a^{-2})\bigr)+\bigl(m(t) a^{-1}\bigr)
 \Bigr)\ .\nonumber
 \end{eqnarray}
 \underbar {Third order with conformal weight $-5$.}
\begin{eqnarray} \edth\edth \thorn' \Phi  &=&
 -6 a^{-5}(n a^{-3/2}),\nonumber \\ 
 \edth \thorn' \thorn' \Phi
 &=& a^{-5}({1\over \sqrt{2}})    \Bigl( 2({b a^{-1}})^2 +15 (n a^{-3/2})^2-4(s(t)
a^{-2})\nonumber\\
&&+3(m(t) a^{-1}) -  i {(b
a^{-1})}\Bigr),\nonumber \\ 
\thorn'  \thorn' \thorn'  \Phi
  &=&  a^{-5} \Bigl(-4(n a^{-3/2})-5(n a^{-3/2})({b a^{-1}})^2 -15(n
 a^{-3/2})^3 \nonumber\\
 &&-9(n a^{-3/2}) (m(t) a^{-1})+ 10(n a^{-3/2})(s(t)a^{-2}) +
  ({b a^{-1}})(e(t) a^{-3/2})\nonumber\\
  &&+(m'(t) a^{-3/2})
 -
(s'(t)a^{-5/2}) \Bigr).
\end{eqnarray}
\underbar {Fourth order with conformal weight $-6$.}
\begin{eqnarray}  \thorn' { X}     &=&    a^{-6}\Bigl((-s''(t) a^{-3})+(m''(t)a^{-2}) +4(n a^{-3/2}) (m'(t)
a^{-3/2}) \nonumber\\
&&-5 (n a^{-3/2}) (s'(t) a^{-5/2}) + ({b a^{-1}})(e'(t)a^{-2})
\nonumber\\
&& -5({b a^{-1}})(n a^{-3/2})(e(t) a^{-3/2}) +(e(t) a^{-3/2})^2\Bigr)
.\ \label{K4h}
\end{eqnarray}

The above expressions have been organised so that, when they are
raised to their respective inverse conformal weights, their first
terms (which will then each be simply $a$, e.g., for second order invariants $(a^{-4})^{-1/4}=a$) on all the right hand
sides  will cancel when ratios are taken.\footnote{The minor corrections in the numerical coefficients in the GHP scalar invariants compared to the published version \cite{BER2}, do not effect any arguments in this section.}

Since the coordinates
$n,b$ are rescaled consistently as $(n a^{-3/2}),({b a^{-1}})$,
respectively, then the nature of the ratios will only depend on
the nature of the  three  functions. This means that all the
expressions $(s(t)a^{-2}),(m(t) a^{-1}), (e(t) a^{-3/2})$ and
$(s'(t)a^{-5/2}), (m'(t) a^{-3/2}),  (e'(t)a^{-2})$ and $(s''(t)
a^{-3}), (m''(t)a^{-2})$ must each be functions of the same {\it
one} variable, and of zero conformal weight. Of course this cannot happen for the generic metric,
but can happen for special cases of the three functions.

For $(s(t)a^{-2}),(s'(t)a^{-5/2}), (s''(t)
a^{-3})$  to be functionally dependent on one variable, it is necessary that $s(t)= s_1t^{-4}$,
where $s_1$ is constant, and similarly it is necessary that $e(t)= e_1t^{-3}$, and  $m(t)= m_1t^{-2}$,
where $e_1, m_1$ are constants;  in all cases we have functions of only the one variable $(at^{2})$. Hence the metric (\ref{metric12}) will admit a homothetic Killing vector for the special case,
when
\begin{eqnarray}
\hbox{d} s^2 &=&\Bigl(a\bigl(2s_1t^{-4}-2am_1t^{-2}-a^2-b^2\bigr)
-e_1^2t^{-6}- n^2\Bigr) \hbox{d} t^2-2a \hbox{d} t \hbox{d} n \nonumber\\
&&+2 n\hbox{d} t \hbox{d} a + 2 e_1t^{-3} \hbox{d} t  \hbox{d} b -
\hbox{d} a^2 - \hbox{d} b^2 \label{metric12h}
\end{eqnarray}
We have already noted that a Killing vector is present when all three functions are constant,
so we can make the further deduction of the presence of {\it both} a Killing and a homothetic Killing vector,
 i.e., {\it  a two-dimensional homothety},
when all three functions are zero,
and
\begin{eqnarray}
\hbox{d} s^2 =\Bigl(-a\bigl(a^2+b^2\bigr)-
n^2\Bigr) \hbox{d} t^2-2a \hbox{d} t \hbox{d} n + 2 n \hbox{d} t
\hbox{d} a  - \hbox{d} a^2 -
\hbox{d} b^2 \label{metric12h0}
\end{eqnarray}

\smallskip

In \cite{ks} it has also been shown that  when there is a
homothetic Killing vector $\bfvartheta$ present,  all intrinsic  GHP
scalars of zero spin and boost weight,  $\eta$ will satisfy the
following condition,
\begin{eqnarray}\vartheta^i  \eta_{,i} = w  \sigma  \eta \label{hcond}
\end{eqnarray}
where $\sigma$ is the constant homothetic  parameter, and  $w$ is the conformal weight of $\eta$.

As noted in the last subsection, we are using an intrinsic GHP tetrad, and  the GHP Cartan invariants
are intrinsic GHP scalars.  For the subclass of spacetimes  with a homothetic Killing vector (\ref{metric12h}) all four  essential base coordinates $a,b,n,t$   are directly identified with GHP Cartan
invariants, and since they also have zero spin and boost weight, they must all satisfy condition (\ref{hcond}) with respect to their respective conformal weights which are easily deduced from the analysis above;
hence the homothetic Killing vector is given by
$$\bfvartheta = -{1\over 2} t {\partial \over \partial t} + {3\over 2} n {\partial \over \partial n} + a{\partial \over \partial a} + b {\partial \over \partial b}
$$
with the parameter choice   $\sigma = 1$. These
results agree with Barnes in \cite{barnes}.

 \subsection{Killing vectors in Koutras-McIntosh
coordinate version.}

For comparison, we can investigate the Killing vector in the version of the
Edgar-Ludwig spacetime in  Koutras-McIntosh coordinates. There are
{\it four} essential base coordinates in all of the cases except
Case {\bf (C(ii))}, where there are only three; hence it follows
immediately that only in Case {\bf (C(ii))} does a Killing vector
exist.  The conditions therefore for a Killing vector in this
spacetime are the three differential equations for $f, g, h$
respectively
$$(ff_{uu}-5f_u^2/4)/f^{-3}= f_0,
$$
$$h(u)-g(u)^2/4 =h_0
$$
$$f^{-1/2}g_{u}= g_0
$$
whose  respective integrals are in the text above, (\ref{intf}), (\ref{inth}), (\ref{intG}).

The actual expression for the Killing vector in these coordinates
will be quite complicated, and would need to be worked out by
directly integrating the Killing equations.

Skea \cite{skea1}, following Koutras \cite{kou}, investigated the
presence of Killing vectors in  Koutras-McIntosh coordinates by
imposing functional dependence on the four essential base
coordinates $\xi_1,\xi_2,\xi_3,\xi_4$ by
$${\partial(\xi_1,\xi_2,\xi_3,\xi_4)\over \partial(u,w,y,z)}=0
$$
and repeating this for different sets of essential coordinates. This
method resulted in a set of three differential equations
\cite{skea1} ), each of which is the derivative of one of the
equations just quoted. The results, by this method, in \cite{skea1}
agree with the results on Killing vectors quoted above, but the
calculations are much longer.

\section{Summary and Discussion}

This paper  highlights the advantages of the
 version of a spacetime which has been derived in GIF (and so
given, as far as possible, in intrinsic coordinates) (\ref{metric12}), compared
with other more familiar versions. Although the spacetime
under consideration here is  part of the familiar
 Kundt family \cite{kundt1}, \cite{kundt2} in the  Ostv\'ath-Robinson-R\'ozga form \cite{ozs} (as
 recently shown directly by Podolsk\'y and Prikryl \cite{pod2}),
 we demonstrate how the intrinsic coordinate
version gives a different insight into the structure of these
spaces.

The simplicity and transparency of this GIF version
(\ref{metric12}), combined with the fact that we are able to carry
it out by hand, gives us a clear unambiguous overview of the
invariant classification of  this class of metrics. Since the
invariant classification procedure is not fully algorithmic, simpler
and more transparent  calculations give important checks, as well as
preventing us overlooking subtle properties. The results obtained in
Section 2 have some minor, but subtle and interesting, disagreements
with the conclusions by Edgar and Vickers \cite{edvic}, and by Skea
\cite{skea1}. The full details of the traditional CLASSI analysis of
the spacetime in the intrinsic coordinates carried out in Section 3
confirms the results in Section 2.

In addition, using this version of the spacetime  (\ref{metric12}), we were able to  obtain trivially
the Killing vector properties, as well as the
 homothetic Killing vector properties by a
straightforward application of the Koutras-Skea algorithm \cite{ks}.
These simple and efficient calculations are in comparison to the
complicated calculations required for the more familiar forms of the
spacetime.

 The motivation to rework Skea's
calculations came from the results of these transparent calculations
in Sections 2 and 3. As noted in the Introduction there are
subtleties connected with these spacetimes which have revealed short
comings in the computer packages  \cite{aman}, \cite{pol}, which are
used to carry out the invariant classification procedure; and once
again these spacetimes have revealed subtleties which had not been
fully appreciated earlier. In Section 4, the traditional CLASSI
analysis of the spacetime in Koutras-McIntosh coordinates is carried
out along the same lines as Skea's analysis; the full details of the
results of Section 2 are again confirmed, and the discrepencies with
Skea's results identified and clarified.  Having this analysis
alongside the one in intrinsic coordinates demonstrates the
simplicity of this GIF version in intrinsic coordinates, and the
fewer possibilities for error and misunderstanding.

\

{\bf Acknowledgements}

SBE wishes to thank Officina Mathematica for supporting a visit to
Universidade do Minho and the Department of Mathematics for Science
and Technology for their hospitality.

MPMR wishes to thank Vetenskapsr\aa det (Swedish Research Council)
for supporting a visit to Link\"opings universitet and the
Mathematics Department for their hospitality.

MB wishes to thank the Mathematics Department at Link\"opings universitet
for its hospitality.

\end{document}